\documentclass[12pt,letterpaper]{article} 

\pdfoutput=1

\usepackage[includeheadfoot,
            marginratio={1:1,2:3}, 
            width=412pt, 
            height=688pt,]{geometry}

\usepackage{amsmath}
\usepackage{amsfonts}
\usepackage{amssymb}

\usepackage{dsfont}
\usepackage{graphicx}
\usepackage{color}
\usepackage{arydshln}
\usepackage{multirow}
\usepackage{mathtools}
\usepackage[nosort]{cite}
\usepackage[margin=10pt,font=small]{caption}
\usepackage{subfigure}

\newcommand{\eq}[1]{\begin{equation}
                     \begin{split} #1 \end{split}
                     \end{equation}}

\newcommand{\op}{\hspace{1pt}}

\allowdisplaybreaks[2]
\numberwithin{equation}{section}

\begin{document}

\vspace*{-1.5cm}

\begin{flushright}
  {\small
  DFPD-2013-TH-19
  }
\end{flushright}

\vspace{3cm}


\begin{center}
{\LARGE
T-duality revisited
}
\end{center}


\vspace{0.4cm}

\begin{center}
  Erik Plauschinn
\end{center}


\vspace{0.4cm}

\begin{center} 
\emph{Dipartimento di Fisica e Astronomia ``Galileo Galilei'' \\
Universit\`a  di Padova \\ Via Marzolo 8, 35131 Padova, Italy}  \\[0.1cm] 
\vspace{0.4cm}
and \\[0.1cm]
\vspace{0.4cm}
\emph{INFN, Sezione di Padova \\
Via Marzolo 8, 35131 Padova, Italy}  \\
\end{center} 

\vspace{2.1cm}


\begin{abstract}
\noindent
We revisit the transformation rules of the metric and Kalb-Ramond field under T-duality, 
and express the corresponding relations in terms of the metric $G$ and the field strength 
$H=dB$. In the course of the derivation, 
we find an explanation for potential reductions of the isometry group in the dual background.

\noindent
The formalism employed in this paper is illustrated with examples based on tori and spheres, where for
the latter we construct a new non-geometric background.
\end{abstract}

\clearpage


\tableofcontents


\section{Introduction}

One of the intriguing features of string theory is that it comprises a rich structure of dualities.
Amongst them is target-space duality, or T-duality in short, which
in its simplest form states that string theory compactified on a 
circle of radius $R$ is equivalent to a compactification on a circle with radius $\alpha'/R$ 
(see  \cite{Giveon:1994fu} and the references therein).
For the circle, the duality group is given by $\mathbb Z_2$, but 
for toroidal backgrounds 
one finds that it takes the form  $O(d,d,\mathbb Z)$ where $d$ denotes 
the number of compact dimensions.
Other dualities for string theory are $S$-duality, which is a strong-weak 
duality, the combination of it with T-duality into so-called $U$-duality, and
also the AdS/CFT duality. 
However, in this work we are primarily interested in T-duality.

Dualities  have played an important role in understanding the structure of string theory
and in uncovering new features; a prominent example thereof is the discovery of D-branes.
More recently, T-duality has helped to find new solutions of the theory with 
surprising and unusual properties. In particular, it turns out that not only ordinary geometric spaces
are eligible backgrounds for string theory, but 
that so-called non-geometric configurations with non-commutative and even 
non-associative features 
are possible as well.
The latter have been found by applying successive T-duality transformations
to a three-torus with non-vanishing field strength $H=dB$ for the Kalb-Ramond field $B$, 
which we are going to review briefly in the following. 
\begin{itemize}

\item Starting from a flat three-torus with non-trivial flux $H$, one performs 
a T-duality transformation 
along a direction of isometry. 
This results in a  twisted torus \cite{Scherk:1978ta,Scherk:1979zr}
with vanishing field strength, where the topology is characterized  by a so-called 
geometric flux $f$ \cite{Dasgupta:1999ss,Kachru:2002sk}.

\item For the twisted torus, an additional T-duality 
transformation
can be performed. The resulting background allows for a locally-geometric description, 
but is globally non-geometric \cite{Hellerman:2002ax}. 
The latter means that when considering a covering of the torus by open neighborhoods, the 
transition functions on the overlap of these charts are not solely given by diffeomorphisms,
and hence such a manifold cannot be described by Riemannian geometry.
However,  if in addition to diffeomorphisms 
one considers T-duality transformations as transition maps \cite{Dabholkar:2002sy},
this space can be globally defined.
This construction is usually called a T-fold \cite{Hull:2004in}, and carries a  so-called $Q$-flux \cite{Shelton:2005cf}.

Note that the $Q$-flux is related to non-commutative features of this background, which have been studied in
a variety of publications 
\cite{Mathai:2004qq,Mathai:2004qc,Grange:2006es,Lust:2010iy,Lust:2012fp,Condeescu:2012sp,Chatzistavrakidis:2012qj,Andriot:2012vb,Bakas:2013jwa}.

\item Finally, for the T-fold it has been argued that  one can formally perform a third 
T-duality transformation \cite{Shelton:2005cf}.
Here it has been found  that the resulting $R$-flux background is not even locally geometric, 
and that it carries a  non-associative structure.

These backgrounds have first been studied from a  mathematical point of view in 
\cite{Bouwknegt:2004ap,Bouwknegt:2004tr}, later in \cite{Ellwood:2006my}, and have
recently been reconsidered in a series of papers \cite{Blumenhagen:2010hj,Blumenhagen:2011ph,Blumenhagen:2011yv,Lust:2012fp,Mylonas:2012pg,Plauschinn:2012kd,Chatzistavrakidis:2012qj,Bakas:2013jwa}.

\end{itemize}
The chain of T-duality transformations reviewed here is usually summarized by illustrating how the 
fluxes in the various backgrounds are related. One finds the following schematic picture \cite{Shelton:2005cf}
\eq{
  \label{t_chain}
  H_{xyz} \quad\xleftrightarrow{\;\;\; T_{z}\;\;\;}\quad
   f_{xy}{}^{z} \quad\xleftrightarrow{\;\;\; T_{y}\;\;\;}\quad
  Q_{x}{}^{yz} \quad\xleftrightarrow{\;\;\; T_{x}\;\;\;}\quad
  R^{xyz} \; .
}
However, let us remark that non-geometric backgrounds can not only be obtained by a chain
of T-duality transformations similar to \eqref{t_chain}, but can also be realized
in the context of asymmetric orbifolds. Examples for such constructions can be found for instance in
 \cite{Dabholkar:2002sy,Hellerman:2002ax,Flournoy:2004vn,Flournoy:2005xe,Hellerman:2006tx,Blumenhagen:2011ph,Condeescu:2012sp,Condeescu:2013yma}.

Non-geometric flux backgrounds, in particular the T-fold background mentioned above, 
have been investigated in a number of 
publications over the years. We do not want to mention all the corresponding references here, but only 
touch upon a few topics.
For one, there are the papers by Hull and collaborators where non-geometric flux
configurations have been studied 
from a doubled-geometry point of view \cite{Hull:2004in,Dabholkar:2005ve,Hull:2006va}.
More recently, non-geometric backgrounds have  been investigated via field redefinitions
for the ten-dimensional supergravity action in 
\cite{Andriot:2011uh,Andriot:2012wx,Andriot:2012an,Blumenhagen:2012nk,Blumenhagen:2012nt,Blumenhagen:2013aia,Andriot:2013xca}.
However, as was found in \cite{Blumenhagen:2013aia}, such methods do not allow for a global description
of non-geometric flux backgrounds.
Let us also mention that a discussion of non-geometric backgrounds 
from a world-sheet point of view can be found in \cite{Flournoy:2005xe,Halmagyi:2008dr,Halmagyi:2009te},
and for studies in the context of  double-field theory
we would like to refer the reader to the reviews \cite{Aldazabal:2013sca,Hohm:2013bwa}.

\bigskip
The main motivation for the present paper is that almost all examples for non-geometric flux backgrounds
are constructed by applying T-duality transformations to tori.
Compared to the landscape of string-theory solutions, this is a very restricted family of configurations. 
In addition, it should be noted that, strictly speaking, toroidal backgrounds with fluxes (and constant dilaton) 
do not solve the string-equations of motion. 
Therefore, one should go beyond the family of tori and search
for new examples of non-geometric spaces.

A well-studied class of proper string-theory backgrounds 
with non-trivial $H$-flux is given by 
Wess-Zumino-Witten models \cite{Witten:1983ar,Gepner:1986wi}, which describe strings moving on group manifolds. One of the simplest examples thereof is based on the group $SU(2)$ and corresponds to  
$S^3$ with a non-vanishing flux $H$.
For such a background, T-duality has been studied for instance in \cite{Bouwknegt:2003vb}, where the authors
found that the dual space is given by a circle bundle over $S^2$. 
However, no indications of a non-geometric structure
have been observed.
In the present paper, we consider a slightly modified setting and investigate 
$S^3\times S^1$ together with a particular choice of $H$-flux. As we describe in detail,
after applying two T-duality transformations to this space, we  arrive at a background which is a 
non-geometric T-fold, and which therefore provides a second class of examples for studying  non-geometry.

\bigskip
In the remainder of this paper, we first review
the sigma-model action for the closed string in section~\ref{sec_sigma}. We study its symmetry structure 
and find that it can be described in terms of the so-called $H$-twisted Courant bracket.
In section~\ref{sec_t-duality}, we follow  Buscher's procedure
\cite{Buscher:1985kb,Buscher:1987sk,Buscher:1987qj}
and gauge a symmetry of the closed-string sigma-model action. Upon integrating out the gauge field, we obtain a formulation
describing the T-dual background. However, in contrast to the Buscher rules which are expressed
in terms of the metric $G$ and the Kalb-Ramond field $B$, here we derive formulas 
for $G$ and $H=dB$. The advantage of this formalism is that no ambiguity in choosing a gauge for the initial
configuration arises.
In section~\ref{sec_tori} we illustrate this method,  and discuss T-duality transformations for tori.
Due to the formulation in terms of the field strength, we are able to obtain generalizations of the 
examples known in the literature. 
In section~\ref{sec_spheres}
we turn to configurations based on spheres, where we first review and generalize
the known results  for the three-sphere, and
subsequently present a new example of a non-geometric background. Finally, 
in section~\ref{sec_sc}, we close with a summary  and conclusion.


\section{Sigma-model action for the closed string}
\label{sec_sigma}

We begin our discussion of T-duality transformations by reviewing
the sigma-model action for the closed string. This action encodes the dynamics of a target-space
metric $G$, an anti-symmetric Kalb-Ramond field $B$ and a dilaton $\phi$, 
and  is usually defined on a compact two-dimensional world-sheet without boundaries. 
However, when considering non-trivial field strengths $H=d B\neq0$, 
one should employ the Wess-Zumino term for the Kalb-Ramond field in order to have a 
globally well-defined description 
on the target space.


\subsubsection*{Action}

Since here we are indeed interested in configurations with $H\neq 0$, 
we formulate the action  in the following way.
Denoting  by $\Sigma$ a three-dimensional Euclidean manifold with compact boundary $\partial \Sigma$,
we have
\eq{
  \label{action_01}
  \mathcal S =& -\frac{1}{4\pi \alpha'} \int_{\partial\Sigma} 
  \Bigl[ G_{ij} \, d X^i\wedge\star d X^j
  + \alpha'R\, \phi \star 1 \Bigr] \\[2mm]
  &-\frac{i}{2\pi \alpha'} \int_{\Sigma} \tfrac{1}{3!}\, H_{ijk}\op dX^i\wedge dX^j\wedge dX^k
  \,.
}
The Hodge-star operator $\star$ is defined on the two-dimensional world-sheet $\partial\Sigma$,  
and the differential is understood as $dX^i(\sigma^{\mu}) = \partial_{\mu} X^i d\sigma^{\mu}$ 
with $\{\sigma^{\mu}\}$ coordinates on $\partial \Sigma$ and on $\Sigma$.
The indices $i,j$ take values $i,j\in\{ 1,\ldots, d\}$ with $d$ the dimension of the target space,
and  $R$ denotes the curvature scalar corresponding to the metric on $\partial\Sigma$.

Note that the choice of a three-manifold $\Sigma$ for a given boundary $\partial\Sigma$ is not unique. 
But, if the (expectation value of the) field strength $H$ is quantized, then the path integral only depends 
on the data of the two-dimensional theory \cite{Witten:1983tw}. 
In our conventions, the quantization condition  reads
\eq{
  \label{quantization}
  \frac{1}{2\pi\alpha'} \int_{\Sigma} H \;\in\; 2\pi \op\mathbb Z \,.
}
Coming to  a slightly more technical point, 
in the following we require the world-sheet fields $X^i$ appearing in the action \eqref{action_01} -- and therefore
also $G_{ij}(X)$, $\phi(X)$ and $H_{ijk}(X)$ --  to be 
well-defined on $\Sigma$ and $\partial\Sigma$. 
More concretely, in order to apply Stoke's theorem in our computations below, 
the scalars $X^i$ should be single-valued as world-sheet fields.
This means that we ignore contributions from winding and momentum modes,
and hence  work in a supergravity approximation.\footnote{Of course, 
if the metric, field strength and dilaton do not depend on $X^i$, only the differentials $dX^i$ appear in the action
\eqref{action_01} and hence winding and momentum modes can 
be incorporated.} Furthermore, we study the sigma-model action \eqref{action_01} at the classical
level, and therefore do not take into account restrictions coming for instance from the vanishing of the conformal anomaly.


\subsubsection*{Invariance under field redefinitions}

In addition to the usual world-sheet symmetries, the action \eqref{action_01} 
can be invariant under field redefinitions of the target-space coordinates $X^i$. Let us therefore 
consider the following change of coordinates for a constant parameter $\epsilon$
\eq{
  \label{iso_trafo_01}
  \delta_{\epsilon} X^i = \epsilon\op k^i(X) \,.
}
Under this variation, the action \eqref{action_01} changes as
\eq{
 \label{variation_09}
  \delta_{\epsilon} \op\mathcal S =&-\frac{1}{4\pi \alpha'} \int_{\partial\Sigma} 
  \, \epsilon\op \Bigl[ \, \mathcal L_k G 
  + \alpha' R \bigl( \mathcal L_k \phi\bigr)\star 1\, \Bigr]
  -\frac{i}{2\pi \alpha'} \int_{\Sigma} \epsilon\, \mathcal L_k H\,,
}
where $\mathcal L_k=d\circ \iota_k + \iota_k\circ d$ 
denotes the Lie derivative  in the direction of the vector $k$,
and $\iota_k$ is the insertion map. 
The variation of the action \eqref{variation_09} vanishes if three conditions are met.
First, $k$ is a Killing vector of the target-space metric $G$, i.e.
\eq{
 \label{iso_01}
  \mathcal L_k \op G = 0 \,,
}  
where we employ the coordinate-free notation $G=G_{ij}\op dX^i\wedge \star dX^j$.
Second, the term involving $\mathcal L_k H$ vanishes, which can be expressed as
\eq{
 \label{iso_02}
  \iota_k H = d v \,,
}
for $v$ a one-form on the target space  \cite{Hull:1989jk,Hull:1990ms} (see also \cite{Belov:2007qj}). Note that  
$v$ in \eqref{iso_02} is defined only up to a closed part.
Finally, the third condition for the variation \eqref{variation_09} to vanish is that
\eq{ 
  \mathcal L_k \phi = k^i \partial_i \phi = 0 \,.
}


\subsubsection*{Algebraic structure of field redefinitions}

Let us now  investigate a situation in which the metric $G$ has several Killing vectors. 
From the commutator property of the Lie derivative it follows that if $k_{(1)}$ and 
$k_{(2)}$ are Killing vectors, also the Lie bracket $k_{(3)}=[k_{(1)},k_{(2)}]_{\rm L}$ is a Killing vector
\eq{
  \label{iso_cond_10}
  \mathcal L_{k_{(3)}} G=\mathcal L_{[k_{(1)},k_{(2)}]_{\rm L}} \op G = 0 \,.
}
This can alternatively be inferred from the closure of the algebra generated by \eqref{iso_trafo_01}.
A similar analysis can be applied to the condition  \eqref{iso_02}. Up to exact terms, 
we find that the one-form $v_{(3)}$ corresponding to the Killing vector $k_{(3)}$ is given by
\eq{
  \label{v3}
   v_{(3)}  = \tfrac{1}{2} \left( \mathcal L_{k_{(1)}} v_{(2)} - \mathcal L_{k_{(2)}} v_{(1)}  \right).
}
The structure describing the algebra of field redefinitions \eqref{iso_trafo_01} which leave
the action \eqref{action_01} invariant can then be formulated using the so-called 
$H$-twisted Courant bracket \cite{Alekseev:2004np}.
The latter is defined as follows
\begin{align}
  \nonumber
  &\bigl[ k_{(1)} + v_{(1)} , k_{(2)} + v_{(2)} \bigr]^H_{\rm C} \\[5pt]
  \nonumber
  &\hspace{30pt}= 
  \bigl[ k_{(1)}  , k_{(2)}  \bigr]_{\rm L} 
   + \mathcal L_{k_{(1)}} v_{(2)} - \mathcal L_{k_{(2)}} v_{(1)} 
   -\tfrac{1}{2} \, d\bigl( \iota_{k_{(1)}} v_{(2)} - \iota_{k_{(2)}} v_{(1)} \bigr)
   -\iota_{k_{(1)}} \iota_{k_{(2)}} H \\[5pt]
  \label{courant_01}
  &\hspace{30pt}= 
   k_{(3)} + v_{(3)}   
   \,,
\end{align}
where the formal sum of a vector and a one-form is considered as an element of the 
generalized tangent space $TM\oplus T^*M$, with $M$ the target-space 
manifold.\footnote{For more details on generalized geometry
we would like to refer the reader to the original papers \cite{Hitchin:2004ut} and \cite{Gualtieri:2003dx},
and for instance to \cite{Grana:2008yw} for a less mathematical discussion.}
Let us make the following remarks:
\begin{itemize}

\item The one-forms $v_{(a)}$ in equation \eqref{iso_02} are defined only up to
terms which are closed, and hence also the Courant bracket \eqref{courant_01} 
is defined only up to closed expressions.
This, in turn,  leaves some freedom to express the algebraic structure of field redefinitions;
for different formulations and further details see for instance \cite{Hull:2009zb}.

\item The automorphisms of the Courant bracket \cite{Gualtieri:2003dx} are 
given by diffeomorphisms of the target-space coordinates, and by transformations involving
a closed two-form $\Omega$.\footnote{The $\Omega$-transform 
introduced here is usually called a $B$-transform. But in order to avoid confusion 
with the Kalb-Ramond field introduced above,  we employ the notation $\Omega$.}
This is also true for the $H$-twisted Courant bracket \eqref{courant_01},
and the explicit form of the $\Omega$-transform reads
\eq{
  k+ v \;\to\; k + (v + \iota_k \Omega)  \hspace{30pt}\mbox{for}\hspace{30pt} d\Omega = 0 \,.
}

\item It is known that 
for the untwisted Courant bracket,
in general the Jacobi identity is not satisfied 
\cite{Gualtieri:2003dx}. In particular, the Jacobiator 
\eq{
{\rm Jac}(A,B,C)_{\rm C}=\bigl[A,[B,C]_{\rm C}\bigr]_{\rm C}+{\rm cyclic}
}
is equal to an exact term. 
In the present context, this does not pose a problem 
since the bracket \eqref{courant_01}
is defined only up to closed expressions (see also \cite{Hull:2009zb}, especially sections 6 and 7). 
For  the $H$-twisted Courant bracket the situation is analogous, 
provided that $H$ satisfies the Bianchi identity $dH=0$ \cite{Gualtieri:2003dx}.

\item As we have shown here, the $H$-twisted Courant bracket \eqref{courant_01}
originates naturally from the global symmetry structure of the sigma model \eqref{action_01}.
The untwisted Courant bracket appears  in the framework of ordinary generalized 
geometry and is well-studied, however, to our knowledge an $H$-twisted version of generalized geometry 
has only been investigated in detail in the mathematical literature \cite{Severa:2001qm}.
But, it would be interesting to apply the $H$-twisted formalism to questions in a 
physical context.

\end{itemize}


\section{T-duality}
\label{sec_t-duality}

We now turn to our discussion of T-duality  and review the transformation rules for the metric, 
Kalb-Ramond field and dilaton.  
These rules can be derived by gauging a target-space isometry in a corresponding  sigma model \cite{Buscher:1985kb,Buscher:1987sk,Buscher:1987qj}, 
which in addition to the original papers has  been discussed in a variety of publications in the past.
The work which is of particular relevance for our approach here can be found in
\cite{Hull:1989jk,Jack:1989ne,Hull:1990ms,Rocek:1991ps,Alvarez:1993qi,FigueroaO'Farrill:1994ns}.


\subsection{Gauging a symmetry}

Let us start by gauging the symmetry \eqref{iso_trafo_01} of the action \eqref{action_01}. 
In the literature, slightly different ways of obtaining a gauged action can be found; 
here we follow the procedure explained  in \cite{Alvarez:1993qi}, which is less restrictive
than the formalism discussed for instance in \cite{Hull:1989jk,Hull:1990ms}.


\subsubsection*{The gauged action}
\label{sec_gauging}

We gauge an isometry of the target-space metric 
by allowing $\epsilon$ in \eqref{iso_trafo_01} to have a non-trivial dependence 
on the world-sheet coordinates.  This implies that we have to introduce a gauge field $A$ and 
replace $dX^i\to dX^i + k^i A$ for the term
involving the metric. For the Wess-Zumino term we keep $dX^i$ unchanged, but introduce an
additional  scalar field $\chi$.
The resulting gauge-invariant action then takes the following form
\begin{align}
  \nonumber
  \widehat{\mathcal S} =&-\frac{1}{4\pi\alpha'} \int_{\partial\Sigma} \Bigl[  
  G_{ij}  (dX^i + k^i A)\wedge\star(dX^j + k^j A)  
  +2\op i (v+d\chi)\wedge A
  + \alpha'R\, \phi \star 1 \Bigr] \\[2mm]
  \label{action_02}
  &-\frac{i}{2\pi \alpha'} \int_{\Sigma}  \tfrac{1}{3!}\, H_{ijk}\op dX^i\wedge dX^j\wedge dX^k\,,
\end{align}
where $k$ denotes the Killing vector of the target-space isometry which has been gauged. 
The explicit form of the symmetry transformations for the fields in the action reads
\eq{
  \label{iso_trafo_02}
  \hat\delta_{\epsilon} X^i = \epsilon\op k^i(X) \,, \hspace{50pt} 
  \hat\delta_{\epsilon} A = - d \op\epsilon\,, \hspace{50pt}
  \hat\delta_{\epsilon} \chi = - \epsilon\op k^i v_i \,,
}
and the corresponding variation of  \eqref{action_02} becomes
\eq{
  \label{variation_02}
  \hat\delta_{\epsilon}\op  \widehat{\mathcal S} = \frac{i}{2\pi \alpha'} \int_{\partial\Sigma}   d\chi \wedge d \epsilon 
   \,,
}
which at this point is non-vanishing.
As mentioned below equation \eqref{quantization}, in the above computations  we have assumed the fields $X^i$ to be single valued on the world-sheet.
However, there can be large gauge transformations if the two-dimensional world-sheet $\partial\Sigma$ is compact  and hence $\epsilon$ can be multivalued \cite{Alvarez:1993qi}. 
Denoting then by  $\gamma$ a non-trivial one-cycle of $\partial\Sigma$ and by $\ell_{\rm s}=2\pi \sqrt{\alpha'}$ the string 
length, we can normalize
\eq{
  \label{period_eps}
  \int_{\gamma} d \epsilon \;\in\; \ell_{\rm s} \op \mathbb{Z} \,.
}
Coming back to \eqref{variation_02}, this variation  vanishes in the path integral
if $\delta_{\epsilon} \op\widehat{\mathcal S}$  is a multiple of $2\pi \op i$. 
Therefore, taking into account \eqref{period_eps}, the scalar field $\chi$ should 
be periodic on the world-sheet $\partial\Sigma$ with period determined by
\eq{
  \label{period_chi}
  \int_{\gamma} d \chi \;\in\; \ell_{\rm s} \op \mathbb{Z} \,.
}


\subsubsection*{Remarks}

After having shown the gauged action \eqref{action_02} to be invariant under the 
transformations \eqref{iso_trafo_02}, let us make the following three remarks.
\begin{itemize}

\item The equations of motion for the scalar field $\chi$ imply that  the field strength $F=dA$
 has to vanish.  
Recalling then from \eqref{period_chi} that $\chi$ can be multivalued,
one can argue that in the path integral the term $d\chi\wedge A$ 
produces a delta-function which sets to zero the holonomy of the gauge field \cite{Rocek:1991ps}.
This implies that $A$ is pure gauge
on the world-sheet $\partial\Sigma$.

\item Note that the ungauged  action \eqref{action_01} is expressed 
in terms of the metric tensor, the dilaton and the field strength $H$,
which are globally defined on the target space.\footnote{Let us mention that we 
call an object {\em globally defined} if it corresponds to a 
single-valued tensor field on the target-space manifold. }
When gauging this action, we implicitly assumed that the Killing vector $k$ is a global object.
But, in \eqref{action_02} also the one-form $v$ appears, which in general is
only  locally defined. Hence, a priori the gauged action \eqref{action_02} is valid only locally
on the target space. 

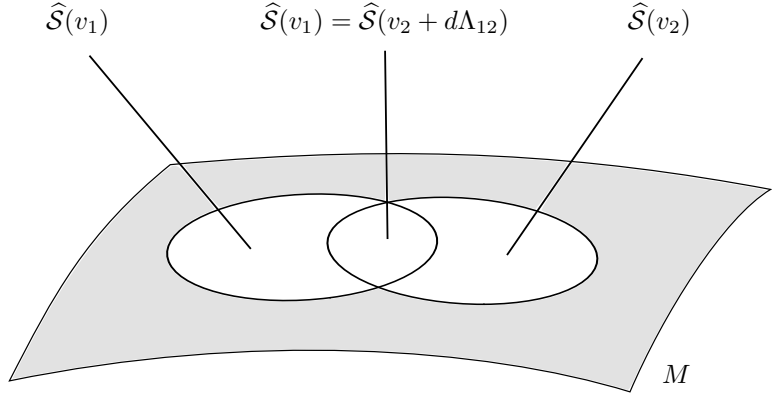
\begin{figure}[t]
\centering
\vspace{10pt}
\begin{picture}(0,0)
\put(103,-127){\footnotesize$ M$}
\put(-130,7){\footnotesize$\widehat{\mathcal S}(v_1)$}
\put(90,7){\footnotesize$\widehat{\mathcal S}(v_2)$}
\put(-49,7){\footnotesize$\widehat{\mathcal S}(v_1)=\widehat{\mathcal S}(v_2+d\Lambda_{12})$}
\end{picture} \\
\includegraphics[width=0.7\textwidth]{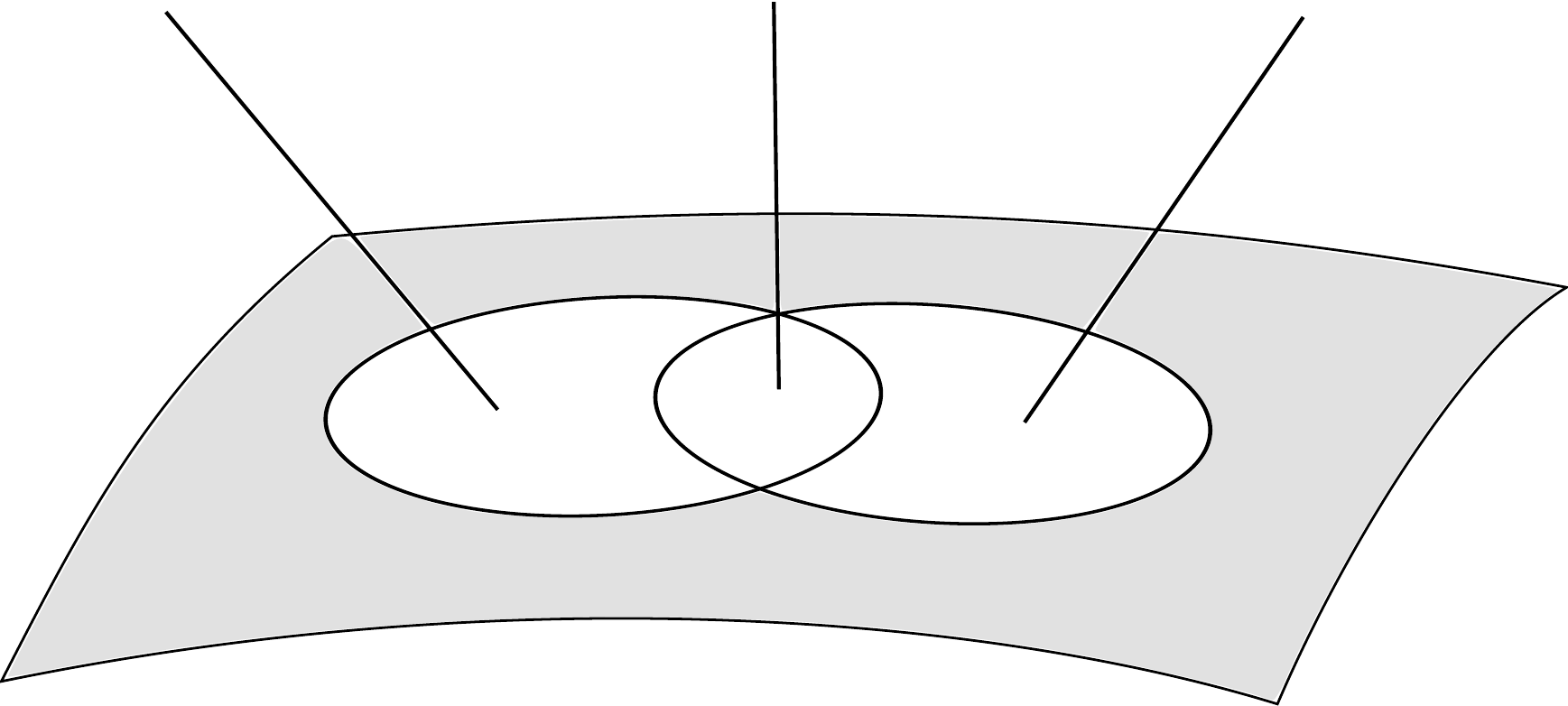} \\
\vspace{10pt}
\caption{Illustration of the local structure of the 
gauged world-sheet action \eqref{action_02} on the target space $M$. The action
depending on $v_1$ is valid in the first patch of  $M$, while
the action depending on $v_2$ is valid in the second patch. On the overlap, the actions agree
provided there exists a function $\Lambda_{12}$ such that $v_1-v_2=d\Lambda_{12}$.
\label{fig_01}}
\end{figure}

However, recall that $v$ is defined up to closed terms, and denote
by $v_{1,2}$ the one-form in two different charts. If on the overlap of these charts 
there exists a function $\Lambda_{12}$ such that $v_{1}-v_2= d\Lambda_{12}$,
one can define $v$, and henceforth the world-sheet action \eqref{action_02}, properly on the whole 
target space. 
This is illustrated in figure~\ref{fig_01}, and has been discussed in detail in \cite{Hull:2006qs}.

\item Note that the assumption of a globally-defined Killing vector $k$ is not always satisfied. 
In this situation, the gauged action \eqref{action_02} is valid only locally on the target space.
This question has also been discussed  in \cite{Hull:2006qs}, and we come back to this point in
sections~\ref{sec_twisted_torus} and~\ref{sec_tw_sphere}.

\end{itemize}


\subsubsection*{Multiple global symmetries}
\label{mult_glob}

Let us now study the situation where the action \eqref{action_01} possesses multiple global symmetries, 
out of which one has been gauged. The Killing vector
corresponding to the gauged symmetry is denoted by $k$, and  Killing vectors for the
 global symmetries are denoted by $Z$. We now want to ask the question under what conditions
the  action \eqref{action_02} is invariant under the remaining global symmetries
\eq{
  \tilde\delta_{\tilde\epsilon} X^i = \tilde \epsilon\op Z^i(X) \,,
}  
where $\tilde \epsilon$ is constant. 
With $[\op\cdot\op,\op\cdot\op]_{\rm L}$ denoting the Lie bracket,
the variation of the gauged action with respect to these global symmetries takes the following form
\eq{
  \tilde\delta_{\tilde\epsilon} \op
  \widehat{\mathcal S} = -\frac{1}{4\pi\alpha'}\int_{\partial\Sigma} \tilde\epsilon\, \Bigl( \:
    [Z,k\op]_{\rm L}^m G_{mi} (dX^i + k^i A)\wedge \star A 
    -2\op i (\mathcal L_Z v) \wedge A
    \:\Bigr) \,.
}
Recall then that the field strength $F=dA$ of the gauge field $A$ vanishes,
and apply Stoke's theorem to the last term in the above expression,
which leads to the condition $d(\mathcal L_Z v)=0$.
Employing furthermore the relation $[\mathcal L_Z,\iota_k]=\iota_{[Z,k]_{\rm L}}$, we find
\eq{
  0 = d(\mathcal L_Z v)=d(\iota_Z\iota_k H) = \iota_{[k,Z]_{\rm L}} H \,.
}
Thus, in order for the gauged action \eqref{action_02} to preserve 
global symmetries of \eqref{action_01}, we have to require
\eq{
  \label{constr_02}
  [Z,k]_{\rm L} = 0 \,.
}
This reflects the familiar statement that when a symmetry 
$H\subset G$ of a symmetry group $G$ is gauged, only the commutant 
$H'=\{ g\in G:hg=gh \;\forall h\in H\}$ remains to be a global symmetry.
We come back to this point in sections~\ref{sec_tfold} and~\ref{sec_sphere}
and illustrate this observation with two examples.


\subsection{Integrating out the gauge field}

The next step in the derivation of the T-duality transformation rules is to integrate  the gauge field out of the 
action \eqref{action_02}. The scalar field $\chi$ can then be interpreted 
as an additional coordinate in an enlarged $(d+1)$-dimensional target space \cite{Rocek:1991ps,Alvarez:1993qi}.


\subsubsection*{World-sheet action for the enlarged target-space}

To integrate out the gauge field $A$,  we determine its equation of motion following from  \eqref{action_02}. After a short computation we find
\eq{
  \label{eom_A}
 |k|^2 A = -  k^i G_{ij}\op dX^j - i \star(v + d\chi) \,,
}
with $|k|^2=k^i G_{ij} k^j$.
If  $|k|^2$  is non-vanishing, we can solve \eqref{eom_A}
for $A$ and substitute the solution back into the action. 
The resulting expression takes the following general form
\eq{
  \label{action_05}
  \check{\mathcal S} =  
  &-\frac{1}{4\pi \alpha'} \int_{\partial\Sigma} \Bigl[ \check G_{\mathsf I\mathsf J} \, 
   d X^{\mathsf I}\wedge\star d X^{\mathsf J}
  + \alpha'R\, \phi \star 1 \Bigr] \\[2mm]
  &-\frac{i}{2\pi \alpha'} \int_{\Sigma} \tfrac{1}{3!}\, \check H_{\mathsf I\mathsf J\mathsf K}\, 
  dX^{\mathsf I}\wedge dX^{\mathsf J}\wedge dX^{\mathsf K} \,,
}
where the indices take values $\mathsf I,\mathsf J,\mathsf K=\{i,\chi\}$ with $i\in\{1,\ldots,d\}$.
Now, to make our following point more clear, let us write the $(d+1)$-dimensional metric 
$\check G$ and field strength $\check H$ in a coordinate-free notation. With  
$\mathsf k = k^i\op G_{ij} \op dx^j$ the one-form dual to the Killing vector $k$, we have
\eq{
  \label{g_and_h_01}
  \check G &= G - \frac{1}{|k|^2}\, \mathsf k \wedge \star \mathsf k +
   \frac{1}{|k|^2}\, (v+ d\chi) \wedge\star (v+ d\chi)\,, \\[3pt]
  \check H &= \frac{1}{|k|^2}\, \iota_k\bigl(\op \mathsf k\wedge H \op\bigr)
  + d\left( \frac{1}{|k|^2}\, \mathsf k \right)\wedge (v+ d\chi)
   \,.
}
From here we see that  $v$ appears explicitly in $\check G$ and $\check H$
which, as explained in section~\ref{sec_sigma}, is not uniquely defined. 
More concretely, $v$  depends on a choice of gauge and does not need to be globally-defined on the target space.
This is in contrast to the usual requirement of the metric and  field strength 
being tensor fields. 
But, as we can  see from \eqref{g_and_h_01},  $v$ always appears in the 
combination $v+d\chi$. This suggests that we should fix the topology of the $(d+1)$-dimensional
space such that 
\eq{
  \label{fix_top}
  e^{\chi} = v + d\chi 
}
is a global one-form.
The latter is characterized by $de^{\chi}= dv = \iota_k H$, which only depends on  $k$ 
and $H$.
Therefore, in the basis  $\{ dX^i, e^{\chi}\}$ both the metric tensor and the field strength
are indeed properly defined. The components of the metric in this basis read 
\eq{
  \label{metric_10}
  &\check G_{IJ} = \left( 
  \begin{array}{cc}
  G_{ij} - \frac{1}{|k|^2}\op k_i k_j& 0 \\[5pt]
  0 & \frac{1}{|k|^2}
  \end{array}
  \right) ,
}
with $I,J=\{i,\chi\}$, and the field strength is given by
\eq{
  \label{field_strength}
  \check H = \frac{1}{|k|^2}\, \iota_k\bigl(\op \mathsf k\wedge H \op\bigr)
  + d\left( \frac{1}{|k|^2}\, \mathsf k \right)\wedge e^{\chi}\,.
}
We also remark that for $|k|^2=0$, the $(d+1)$-dimensional metric and 
field strength become singular. In the following, we therefore exclude this situation. 
(See however \cite{Rocek:1991ps} for a brief discussion of this issue from a conformal field 
theory point of view.)


\subsubsection*{Killing vector}

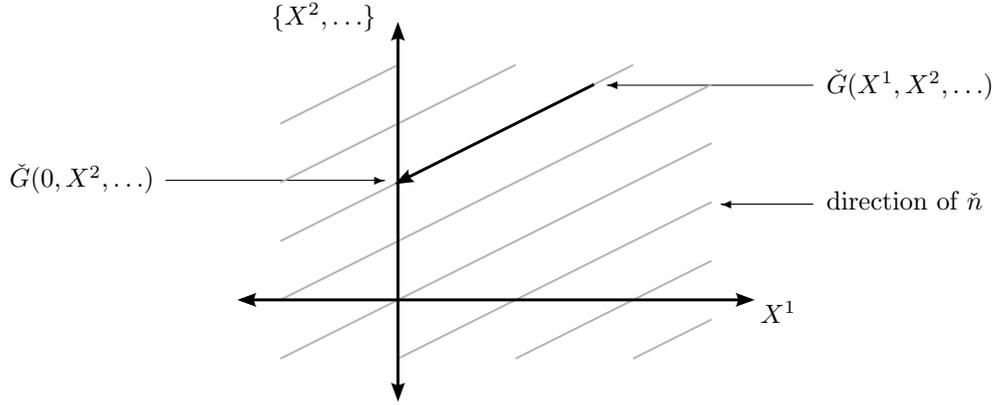
\begin{figure}[t]
\centering
\vspace{10pt}
\begin{picture}(0,0)
\put(100,-115){\footnotesize$X^1$}
\put(-85,-2){\footnotesize$\{ X^2,\ldots\}$}
\put(125,-28){\footnotesize$\check G(X^1,X^2,\ldots)$}
\put(120,-25){\vector(-1,0){78}}
\put(-184,-63.5){\footnotesize$\check G(0,X^2,\ldots)$}
\put(-125,-61){\vector(1,0){83}}
\put(120,-70){\vector(-1,0){35}}
\put(125,-72){\footnotesize direction of $\check n$}
\end{picture} \\
\includegraphics[width=0.475\textwidth]{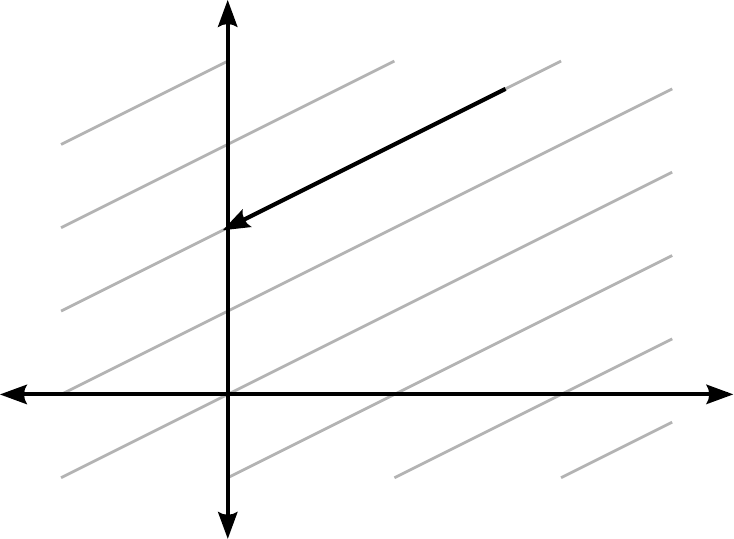} \\
\vspace{5pt}
\caption{Illustration of how the metric $\check G$
can be transported along the Killing vector $\check n$ to $X^1=0$ (for $k^1$ non-vanishing).
Hence, $\check G$ is completely characterized by its values at $X^1=0$. 
Since also the Lie derivative of $\check H$ and $\phi$ along $\check n$ vanishes, the same holds
for these fields.
\label{fig_02}}
\end{figure}

We now determine possible Killing vectors for the $(d+1)$-dimensional 
metric $\check G_{IJ}$  shown in  \eqref{metric_10}.
As one can check explicitly, one Killing vector for $\check G_{IJ}$ is always
given by
\eq{
  \label{vec_01}
  \check n = \binom{\,k^i\,}{0} \,,
}
and hence $\mathcal L_{\check n} \check G=0$.
Furthermore, we note that also the Lie derivative of $\check H$ and of $\Phi$
in the direction of $\check n$ vanishes, 
and  therefore we have
\eq{
  \label{iso_10}
  \check{\mathcal L}_{\check n} \check G = 0\,,\hspace{60pt}
  \check{\mathcal L}_{\check n} \check H = 0\,,\hspace{60pt}
  \check{\mathcal L}_{\check n} \phi = 0
  \,,
}
which implies that  we can move these fields along the direction of the Killing vector
\eqref{vec_01}. More concretely, without loss of generality let us assume that the first component $k^1$ of $\check n$ is non-vanishing
\eq{
  k^1 \neq 0 \,.
}
We can then shift the fields in the
action \eqref{action_05} along the Killing vector \eqref{vec_01} to a convenient 
point in the $(d+1)$-dimensional space, say $X^1=0$,
which we assume  for the following.
This procedure is illustrated in figure~\ref{fig_02}.


\subsubsection*{Adapted coordinates}

Next, we note that  even though the original $d$-dimensional metric $G_{ij}$ is 
usually non-degenerate, 
the matrix $\check G_{IJ}$ in \eqref{metric_10} 
has one vanishing eigenvalue. The corresponding null-eigenvector 
is given \eqref{vec_01}. 
With $k^1$ non-zero, this allows us to perform a change of 
coordinates as follows
\eq{
  \label{coc_01}
  \check{\mathcal G}_{IJ} =  \bigl( \mathcal T^T \check G \,\mathcal T 
  \bigr)_{IJ}\,,
  \hspace{70pt}
  \mathcal T^I{}_J=
  \renewcommand{\arraystretch}{1.35}
  \arraycolsep6pt
  \dashlinedash2pt
  \dashlinegap4pt  
  \left( \begin{array}{c:c@{\hspace{17pt}}c@{\hspace{17pt}}c|c}
  k^1 & \multicolumn{3}{c|}{0} & 0 \\ \hdashline
  k^2 &  &&  & \multirow{3}{*}{0}  \\[-2pt]
  \vdots & & \mathds 1 && \\[-2pt]
  k^d & &&  &  \\ \hline
  0 & \multicolumn{3}{c|}{0}  & 1
  \end{array}
  \right).
}
In the transformed matrix $\check{\mathcal G}_{IJ}$ 
all entries along the $I,J=1$ direction vanish, and 
we therefore arrive at
the  expression
\eq{
  \label{coc_02}
  \check{\mathcal G}_{IJ} =
  \renewcommand{\arraystretch}{1.4}
  \arraycolsep6pt
  \dashlinedash2pt
  \dashlinegap4pt  
  \left( \begin{array}{c:c@{\hspace{2pt}}c@{\hspace{2pt}}c|c}
  0 & \multicolumn{3}{c|}{0} & 0 \\ \hdashline
  &  &&  &   \\[-12pt]
  0 & & \check G_{\alpha\beta} && 0 \\[-12pt]
  & &&  &  \\ \hline
  0 & \multicolumn{3}{c|}{0}  & \check G_{\chi\chi}
  \end{array}
  \right) ,
}
where $\alpha,\beta\in\{2,\ldots,d\}$.
This change of coordinates corresponds to what is sometimes called 
\emph{going to adapted coordinates} in the literature.
Turning then to the $(d+1)$-dimensional field strength \eqref{field_strength} and employing
the matrix $\mathcal T^I{}_J$, we transform $\check H$ as follows
\eq{
  \check{\mathcal H}_{IJK} = \check H_{ABC} \mathcal T^A{}_I\mathcal T^B{}_J\mathcal T^C{}_K\,.
}  
Similarly to the transformed metric $\check{\mathcal G}_{IJ}$, we again find that all 
components of $\check{\mathcal H}$ 
along the $I=1$ direction vanish, that is
\eq{
  \label{coc_13}
  \check{\mathcal H}_{1 JK } = 0 \,.
}

From  \eqref{coc_02} and \eqref{coc_13} one might conclude that in the action 
\eqref{action_05} the $I=1$ direction has dropped out, and that one has effectively 
arrived at a $d$-dimensional formulation.
However, that conclusion is premature. 
To illustrate this point, let us consider the transformed basis of one-forms given 
by $e^I = (\mathcal T^{-1} )^I{}_J \op dx^J$.
For the transformation matrix $\mathcal T$ shown in \eqref{coc_01}, we find 
\eq{
  \label{coc_11}
  e^1 = \frac{1}{k^1}\, dx^1\,, \hspace{50pt}
  e^{\alpha} = dx^{\alpha}- \frac{k^{\alpha}}{k^1}\, dx^1 \,, \hspace{50pt}
  e^{\chi} \equiv e^{\chi} \,,
}
where again $\alpha=2,\ldots, d$.
Note that the algebra of one-forms $\{ e^{\alpha},e^{\chi}\}$ does not close, but 
requires the basis of forms in the full $(d+1)$-dimensional space. More concretely, we have
\eq{
  de^{\alpha} & = 
   \frac{1}{k^1}\op\Bigl(  k^1 \partial_{\beta} k^{\alpha} -k^{\alpha}\partial_{\beta} k^1 \Bigr)\,
  e^1\wedge e^{\beta} \,, \hspace{40pt}
  de^{\chi}  = 
  \frac{1}{2}\op k^m H_{m\alpha\beta} \op e^{\alpha}\wedge e^{\beta} \,.
}
Therefore, in order for $\{ e^{\alpha},e^{\chi}\}$ to close on itself and to properly reduce the $(d+1)$-dimensional 
target space to  $d$ dimensions,
for all $\alpha,\beta\in\{2,\ldots, d\}$ we  have to require 
\eq{ 
  \label{coc_04}
  0= k^1 \partial_{\beta} k^{\alpha} -k^{\alpha}\partial_{\beta} k^1 \,.
 }


\subsection{Transformation rules}
\label{buscher_rules}

At this point, we have performed all important steps in the derivation of the T-duality
transformation rules. However, before presenting the final formulas, 
let us briefly summarize our discussion so far.
\begin{enumerate}

\item In section~\ref{sec_gauging}, we have gauged a symmetry
of the world-sheet action \eqref{action_01}, which 
corresponds to a target-space isometry. 
The resulting gauged action is shown in equation \eqref{action_02}. 

\item Next, we integrated the gauge field out of the action, leading to the
sigma model \eqref{action_05} which
describes a $(d+1)$-dimensional target space with metric $\check G$ and field strength 
$\check H$.

\item In order for $\check G$ and $\check H$ to be well-defined objects, not depending 
on a choice of gauge, we fixed the topology of the $(d+1)$-dimensional space such that 
$e^{\chi} = v + d\chi$ shown in \eqref{fix_top} is globally defined.

\item We also observed that the Lie derivative of $\check G$, $\check H$ and of the dilaton 
along the  vector \eqref{vec_01} vanishes.
With the help of this $(d+1)$-dimensional isometry, 
we have then set the $X^1$-dependence of the above fields to a convenient value, for instance to $X^1=0$
(see figure~\ref{fig_02}).

\item Furthermore, the matrix $\check G_{IJ}$ displayed in \eqref{metric_10} has a null-eigenvector
also given by  \eqref{vec_01}. Employing the latter,
we have performed a change of coordinates leading to 
the metric $\check{\mathcal G}_{IJ}$ and field strength $\check{\mathcal H}_{IJK}$
in which all components along the $I=1$ direction vanish. 
In order for the algebra of one-forms $\{e^2,\ldots, e^d,e^{\chi}\}$
to close, we have  imposed the condition \eqref{coc_04}.

\end{enumerate}
After these steps, we arrived at a situation where the complete $e^1$-dependence in the action 
\eqref{action_05} has disappeared, and where all $X^1$-dependence has been fixed.
This effectively reduces the metric $\check{\mathcal G}$ and field strength $\check{\mathcal H}$
from $(d+1)$ dimensions to $d$ dimensions.


\subsubsection*{Final results}
\label{buscher_page}

The T-duality transformation rules 
for the dual $d$-dimensional metric $\mathcal G$
are then obtained 
by simply deleting the first row and first column in \eqref{coc_02}.
Employing the basis of one-forms $\{e^{\chi},e^{\alpha}\}$
with $\alpha\in\{2,\ldots,d\}$, we have
\eq{
  \mathcal G_{\chi\chi} = \left. \frac{1}{|k|^2}\, \right\vert_{X^1=0} \,,
  \hspace{40pt}
  \mathcal G_{\alpha\chi} = 0 \,,
  \hspace{40pt}
  \mathcal G_{\alpha\beta} =  \left. 
  G_{\alpha\beta} - \frac{k_{\alpha}\op k_{\beta} }{|k|^2}
  \, \right\vert_{X^1=0} \,.
  \label{buscher_1}
}
Let us emphasize that $|k|^2= k^iG_{ij} k^j$ and $k_{\alpha} = G_{\alpha j}k^j$ are computed using 
the original metric $G_{ij}$ with $i,j\in\{1,2,\ldots,d\}$.
The maybe somewhat unfamiliar form of these transformation rules for the metric stems from the fact 
that we use a
basis of globally-defined  one-forms $\{e^{\chi},e^{\alpha}\}$, which is not necessarily closed
\eq{
  \label{buscher_03}
  de^{\alpha} = 0 \,, \hspace{70pt}
  de^{\chi} = \frac{1}{2}\, k^m H_{m\alpha\beta} \,e^{\alpha}\wedge e^{\beta} \,.
}
The components of the dual $d$-dimensional field strength in this basis take the following form
\eq{
  \label{buscher_02}
  \mathcal H_{\alpha\beta\gamma} &= \left.
  H_{\alpha\beta\gamma} - \frac{1}{|k|^2}\Bigl( 
  k_{\alpha} \op k^m H_{m\beta\gamma}
  +k_{\beta} \op k^m H_{m\gamma\alpha}
  +k_{\gamma} \op k^m H_{m\alpha\beta} \Bigr) \, \right\vert_{X^1=0}\,, 
  \\[8pt]
  \mathcal H_{\alpha\beta\chi} &=   \left. \partial_{\alpha}\left( \frac{k_{\beta}}{|k|^2} \right)
  - \partial_{\beta}\left( \frac{k_{\alpha}}{|k|^2} \right) \,\right\vert_{X^1=0} \,.
}
Finally, the transformation properties of the dilaton can be determined from a one-loop computation 
\cite{Buscher:1985kb}. Without giving further details, here we only quote the result for the dual dilaton $\Phi$ as
\eq{
   \Phi = \phi + \frac{1}{2}\log \det \mathcal G \,.
}
In the derivation of the  transformation rules presented here,
we have imposed certain restrictions on the vector $k$.  
These are summarized as follows
\eq{
  \label{restrictions}
  \renewcommand{\arraystretch}{1.4}
  \begin{array}{@{}l@{\hspace{45pt}}l@{}}
  \mbox{Killing vector condition} & 0 = G_{im}\partial_j k^m + G_{jm}\partial_i k^m + k^m \partial_m G_{ij} \,, \\
  \mbox{non-vanishing norm} & |k|^2 = k^i G_{ij} k^j \neq0 \,, \\
  \mbox{T-duality direction} & k^1  \neq0  \,, \\
  \mbox{closure of the algebra} & 0 = \left. k^1 \partial_{\alpha} k^{\beta} - k^{\beta} \partial_{\alpha} k^1 
  \,\right\rvert_{X^1=0}
  \,.
  \end{array}  
}


\subsubsection*{Remarks}

We close this section with the following remarks. 
\begin{itemize}

\item Note that the change of coordinates shown in equation \eqref{coc_01}
is particularly suitable for Killing vectors corresponding to translations, where  one 
component of $k$ is always non-vanishing. 
For Killing vectors describing rotations,  adapted coordinates
are given by  spherical coordinates.

Furthermore, if the Killing vector $k$ is globally defined, one can always employ the Gram-Schmidt process 
to obtain a system of coordinates where $k$ has only one non-vanishing component. 
However, for local Killing vectors this is not possible and the constraint \eqref{coc_01}
is non-trivial.

\item From the equations shown in \eqref{buscher_03}
we infer that the topology of the dual space corresponds to a non-trivial circle bundle.
Indeed, the circle in the $\chi$-direction is non-trivially fibered over the base space 
in the remaining directions. We denote this bundle by $\mathcal E$, and compute its first Chern class as
\eq{
  \label{math_01}
  c_1 (\mathcal E) = \frac{1}{2}\, k^m H_{m\alpha\beta} \,e^{\alpha}\wedge e^{\beta}\,,
}
which is in agreement with results in the mathematical literature on this subject 
\cite{Bouwknegt:2003vb,Bouwknegt:2003wp,Bouwknegt:2003zg}.
In particular, in the latter papers it has been shown that 
\eq{
  \label{math_02}
  c_1(\mathcal E) = \int_{S^1} H \,,
}
where $H$ is the original field strength and $S^1$ is the circle along which the T-duality transformation
is performed. A short computation then shows that the right-hand sides of \eqref{math_01} and
\eqref{math_02} are indeed equal.

\item Above equation \eqref{fix_top}, we argued that in order for the $(d+1)$-dimensional
metric and field strength to be globally defined, the topology of the dual space has to be changed.
We  emphasize that this observation is not new, but has first been made in 
\cite{Alvarez:1993qi,Bouwknegt:2003vb,Bouwknegt:2003wp}.

\end{itemize}


\section[Examples I \hspace{5.35pt}-- Tori]{Examples I -- Tori}
\label{sec_tori}

Let us now illustrate the transformation rules derived above with some examples. 
Here, we focus on tori with non-vanishing field strength $H$, and in the next section we turn to 
spheres.
For the case of the torus, T-duality has been studied extensively in the literature, for which
references can be found in  the introduction.


\subsection{Torus}
\label{ex1_flat_t}

Let us start our discussion by first considering a flat three-torus 
together  with a non-trivial field strength $H$. The metric in the standard basis of 
one-forms $\{dX^1,dX^2,dX^3\}$ is chosen to be of the form
\eq{
  \label{metric_01}
  G_{ij} = \left( \begin{array}{ccc} 
  R_1^2 & 0 & 0 \\ 
  0 & R_2^2 & 0 \\
  0 & 0 & R_3^2
  \end{array} \right),
}  
and the topology is characterized by the identifications $X^i\simeq X^i+\ell_{\rm s}$ for $i=1,2,3$.
The field strength $H=dB$ of the Kalb-Ramond field  is essentially unique in three dimensions
and, taking into account the quantization condition \eqref{quantization}, is given by
\eq{
  H = h \, dX^1 \wedge dX^2\wedge dX^3 \,,
  \hspace{60pt}
  h\in \ell_{\rm s}^{-1}\op \mathbb Z \,.
  \hspace{-30pt}
}


\subsubsection*{Symmetry structure}

For the metric \eqref{metric_01}, there are three linearly independent directions of 
isometry which are  of interest to us.
In a basis $\{\partial_1,\partial_2,\partial_3\}$ dual to the one-forms, they are given by the 
Killing vectors
\eq{
  \label{ex1_killing_02}
  \arraycolsep2pt
  k_{(1)} = \left( \begin{array}{c} 1 \\ 0 \\ 0 \end{array} \right)  ,\hspace{50pt}
  k_{(2)} = \left( \begin{array}{c} 0 \\ 1 \\ 0 \end{array} \right)  ,\hspace{50pt}
  k_{(3)} = \left( \begin{array}{c} 0 \\ 0 \\ 1 \end{array} \right)  .
}  
The one-forms $v_{(a)}$ corresponding to \eqref{ex1_killing_02} are defined by equation \eqref{iso_02},
and up to exact terms they can be written as
\eq{
\arraycolsep1.2pt
\renewcommand{\arraystretch}{1.4}
\begin{array}{lclclcr}
  v_{(1)} &=&  h \op\alpha_1 & X^2 \op dX^3 & - h \op \alpha_2 & X^3 \op dX^2 \,,& \hspace{60pt} 
    \alpha_1 + \alpha_2 = 1 \,, \\
  v_{(2)} &=&  h \op\beta_1 & X^3 \op dX^1 & - h \op \beta_2 & X^1 \op dX^3 \,,& \hspace{60pt} 
    \beta_1 + \beta_2 = 1 \,, \\
  v_{(3)} &=&  h \op\gamma_1 & X^1 \op dX^2 & - h \op \gamma_2 & X^2 \op dX^1 \,,& \hspace{60pt} 
    \gamma_1 + \gamma_2 = 1 \,,
\end{array}    
}
where $\alpha_{m}$, $\beta_{m}$ and $\gamma_{m}$ are constant.
Note that these one-forms are not globally
defined on the torus. However, due to the equivalence 
$v_{(a)} \simeq v_{(a)} + d \Lambda$ for a function $\Lambda$,
we can define the $v_{(a)}$  on local charts and 
cover the torus consistently (see figure~\ref{fig_01} and reference \cite{Hull:2006qs}).

Let us also recall from section~\ref{sec_sigma} that the global symmetry algebra for the sigma model is
given by the $H$-twisted Courant bracket \eqref{courant_01}. Since the Killing vectors
\eqref{ex1_killing_02} commute, the corresponding algebra is trivial, that is up to exact terms
we have
\eq{
  \bigl[ k_{(a)} + v_{(a)} , k_{(b)} + v_{(b)} \bigr]^H_{\rm C} = 0 \,, \hspace{50pt}
  a,b=1,2,3\,.
}


\subsubsection*{T-duality transformation}

We  now perform a T-duality transformation on the above configuration. 
Because we are interested in a general situation, we choose the Killing vector along
which we dualize as
\eq{
  \label{ex1_killing}
  \arraycolsep2pt
  \hspace{50pt}
  k = \left( \begin{array}{c} k^1 \\ k^2 \\ k^3 \end{array} \right)  
  \hspace{100pt}
  \mbox{with~}k^i={\rm const}.
}
Without loss of generality, let us require $k^1\neq0$ so that the restrictions
shown in \eqref{restrictions} are satisfied. 
After applying the transformation rules, a new basis of globally-defined one-forms 
$\{e^{\chi},e^2,e^3\}$ should be used, which is characterized by \eqref{buscher_03}  as
\eq{
  \label{ex1_basis}
  d e^{\chi} = k^1  h\, e^2\wedge e^3 \,, \hspace{60pt}
  d e^2 = 0 \,, \hspace{60pt}
  d e^3 = 0 \,.
}  
Employing this basis, the dual metric is determined  by equation \eqref{buscher_1}
and, with  $i,j\in\{\chi,2,3\}$,  takes the following form  
\eq{
  \label{metric_17}
  \renewcommand{\arraystretch}{1.25}
  \mathcal G_{ij} = \frac{1}{|k|^2} \left( \begin{array}{ccc} 
  1 & 0 & 0 \\ 
  0 & \displaystyle R_2^2 \op\Bigl[\op R_1^2 (k^1)^2 + R_3^2 (k^3)^2 \op \Bigr]
  & \displaystyle - R_2^2 \op R_3^2\op k^2 k^3 \\
  0 & \displaystyle - R_2^2 \op R_3^2\op k^2 k^3 
  & \displaystyle R_3^2 \op\Bigl[\op R_1^2 (k^1)^2 + R_2^2 (k^2)^2 \op \Bigr]
  \end{array} \right).
}  
Note that here we defined
$|k|^2 = R_1^2\op(k^1)^2+R_2^2\op(k^2)^2+R_3^2\op(k^3)^2$, and that
the dual of the field strength is determined from equation  \eqref{buscher_02}  as
\eq{
  \label{ex1_11}
  \mathcal H = 0 \,.
}


\subsubsection*{Choice of a local geometry}

Let us discuss the above background in some more detail.
First, note that the rather complicated-looking expression for the metric \eqref{metric_17} 
simplifies if we choose $k^1=1$ and $k^2=k^3=0$. In that case, we arrive at the familiar form
\eq{
  \label{ex1_matrix12}
  \renewcommand{\arraystretch}{1.2}
  \arraycolsep3pt
  \mathcal G_{ij} =  \left( \begin{array}{ccc} 
  \frac{1}{R_1^2} & 0 & 0 \\ 
  0 & R_2^2 & 0\\
  0 & 0 & R_3^2
  \end{array} \right).
}  
Next, we recall that the matrix in \eqref{ex1_matrix12} is written in the globally-defined 
basis shown in \eqref{ex1_basis}.
However, to get a better geometric understanding let us introduce local coordinates $\{Y^1,Y^2,Y^3\}$ and 
express the basis of one-forms as
\eq{
  \label{ex1_basis17}
  e^{\chi} &= d Y^1 + k^1\op h\op\Bigl( \kappa_1 \op Y^2 dY^3 - \kappa_2 \op Y^3 dY^2 \Bigr)\,,
  \hspace{60pt} \kappa_1+\kappa_2=1 \,, \\
  e^2 &= dY^2 \,, \\[4pt]
  e^3 &=dY^3 \,,
}
where  the freedom of choosing a local system of coordinates is parametrized by the constants
$\kappa_1$ and $\kappa_2$.
In the local basis given by \eqref{ex1_basis17}, the matrix \eqref{ex1_matrix12} then 
takes the following form
\eq{
  \label{ex1_matrix14}
  \renewcommand{\arraystretch}{1.6}
  \arraycolsep10pt
  \widetilde{\mathcal G}_{ij} =  \left( \begin{array}{ccc} 
  \frac{1}{R_1^2} 
  & - \frac{1}{R_1^2}\op k^1 h \op \kappa_2 \op Y^3 
  & + \frac{1}{R_1^2}\op k^1 h \op \kappa_1 \op Y^2 \\ 
  - \frac{1}{R_1^2}\op k^1 h \op \kappa_2 \op Y^3 
  & R_2^2 + \frac{1}{R_1^2} \bigl[ k^1\op h\op \kappa_2 Y^3 \bigr]^2
  & -\frac{1}{R_1^2}\bigl[ k^1 \op h\bigr]^2 \kappa_1\kappa_2 Y^2 Y^3 \\
  + \frac{1}{R_1^2}\op k^1 h \op \kappa_1 \op Y^2 
  & -\frac{1}{R_1^2}\bigl[ k^1 \op h\bigr]^2 \kappa_1\kappa_2 Y^2 Y^3 
  & R_3^2 + \frac{1}{R_1^2} \bigl[ k^1\op h\op \kappa_1 Y^2 \bigr]^2
  \end{array} \right).
}  
In order to investigate this geometry, we consider the limiting case of 
say $\kappa_1=0$ and $\kappa_2=1$. For this particular choice, the metric
\eqref{ex1_matrix14} describes a two-torus in the $\{Y^1,Y^2\}$-direction which is non-trivially 
fibered over a circle in the $Y^3$-direction. 
Indeed, as one can also see from \eqref{ex1_basis17}, 
for a well-defined metric one has to shift
 $Y^1\to Y^1 + k^1 h\op\ell_{\rm s}\op Y^2$
when going around the
circle as $Y^3\to Y^3 + \ell_{\rm s}$. This background is know as a twisted torus \cite{Scherk:1978ta,Scherk:1979zr}.
The other limiting case with $\kappa_1=1$ and $\kappa_2=0$ is on equal footing. Here, one
obtains a torus in the $\{Y^1,Y^3\}$-direction which is non-trivially 
fibered over a circle in the $Y^2$-direction. 
However, for  $\kappa_1\neq0$ and $\kappa_2\neq 0$ the situation is more complicated 
and in general one cannot find a two-torus fibered over a circle. 
In fact, as mentioned around equation \eqref{math_01}, topologically we have a circle bundle in the 
$Y^1$-direction over a  two-torus in the $\{Y^2,Y^3\}$-direction.

Furthermore, let us recall that 
in section~\ref{buscher_rules}
we derived the T-duality  rules for the metric $G$ and the field strength $H=dB$.
In this case, there is no ambiguity of choosing a gauge for $B$ in the initial configuration,
and we can perform a T-duality in any direction along the torus.
However, for the dual background there is a freedom of choosing local coordinates,
parametrized by $\kappa_1$ and $\kappa_2$. 
This is in agreement with the commonly-known Buscher rules for the metric and $B$-field 
\cite{Buscher:1985kb,Buscher:1987sk,Buscher:1987qj}.
Here, one has to first choose a gauge for $B$ which then determines the  dual space. 
However,  note that a choice of gauge, and correspondingly a choice of local coordinates, should
not change the dual geometry. In view of the above example, let us therefore consider  two  local 
systems of coordinates $\{Y^1,Y^2,Y^3\}$ and $\{\widetilde Y^1, \widetilde Y^2, \widetilde Y^3\}$.
These can be related by applying the redefinitions
\eq{
  \widetilde Y^1  = Y^1 + k^1\op h \bigl( \kappa_1 - \widetilde\kappa_1 \bigr)\op Y^2\op Y^3 \,,
  \hspace{40pt}
  \widetilde Y^2  = Y^2 \,,
  \hspace{40pt}
  \widetilde Y^3  = Y^3 \,,
}
and hence the metrics written in these coordinates are diffeomorphic, and so the geometry is the same.


\subsubsection*{Remarks}

We close this section with a discussion of some further questions in relation to the 
dual background.
\begin{itemize}

\item First, let us go back to the metric  \eqref{metric_17} expressed in the basis
of one-forms $\{e^{\chi},e^2,e^3\}$. In contrast to the previous paragraph, we now consider the 
general case where all components of the Killing vector \eqref{ex1_killing} can be non-vanishing.
We then see that the base manifold in the $\{e^2,e^3\}$-direction is a tilted torus, which is illustrated in
figure~\ref{fig_04}. 
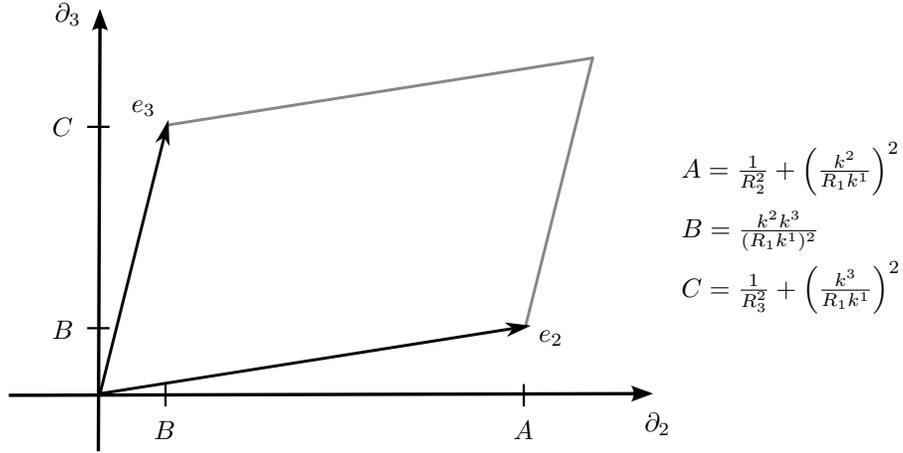
\begin{figure}[t]
\centering
\vspace{10pt}
\begin{picture}(0,0)
\put(66,-161){\footnotesize$\partial_2$}
\put(-157,-6){\footnotesize$\partial_3$}
\put(26,-127){\footnotesize$e_2$}
\put(-128,-40){\footnotesize$e_3$}
\put(16.5,-164){\footnotesize$A$}
\put(-119.5,-164){\footnotesize$B$}
\put(-158,-126){\footnotesize$B$}
\put(-158,-49){\footnotesize$C$}
\put(80,-65){\footnotesize$A= \frac{1}{R_2^2}+\left( \frac{k^2}{R_1 k^1}\right)^2$}
\put(80,-87.5){\footnotesize$B= \frac{k^2 k^3}{(R_1 k^1)^2}$}
\put(80,-110){\footnotesize$C= \frac{1}{R_3^2}+\left( \frac{k^3}{R_1 k^1}\right)^2$}
\end{picture} \\
\includegraphics[width=0.85\textwidth]{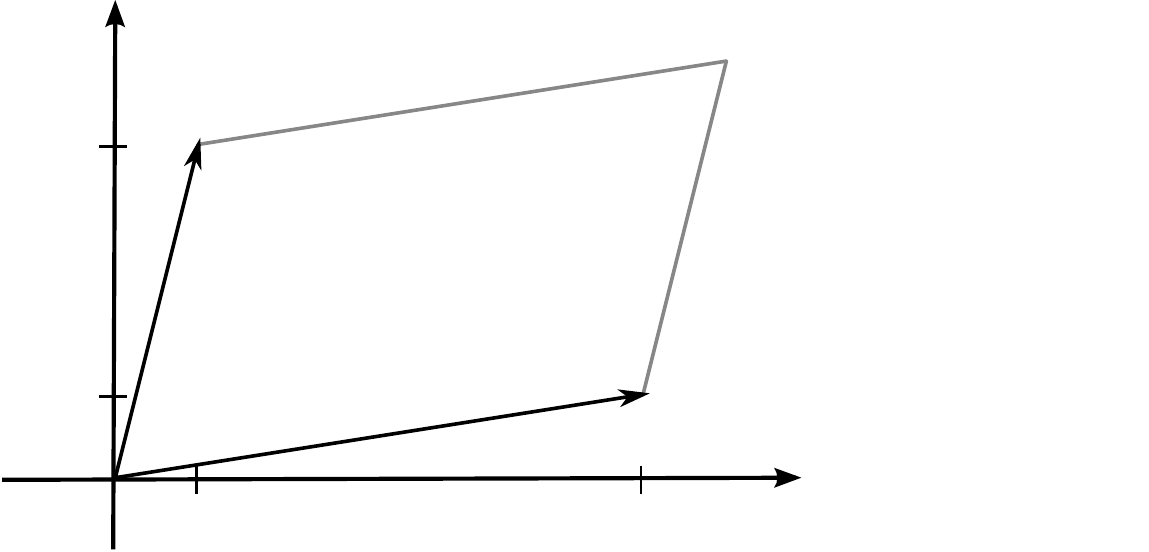} \\
\vspace{5pt}
\caption{Illustration of the tilted torus in the $\{e^2,e^3\}$-direction of the metric \eqref{metric_17}.
The vector fields $e_2$ and $e_3$  dual to the one-forms $e^2$ and $e^3$ have been 
determined via the metric \eqref{metric_17} as $\mathcal G(e_{\alpha},\cdot\op)=e^{\alpha}$.
Note that for $k^2$ or $k^3$ vanishing, the two-torus becomes rectangular.
\label{fig_04}}
\end{figure}
The explicit expressions for the complex structure $\tau$ 
and the volume of this torus are not 
very illuminating, but we give them  for completeness.
Employing the constants introduced in figure~\ref{fig_04}, they read
\eq{
  &\tau = \frac{1}{A^2+B^2} \Bigl[ \bigl( A+C\bigr) B \;+\;i\op \bigl( AC-B^2\bigr) \Bigr] \,, \\[10pt]
  &{\rm vol}(\mathbb T^2) = AC- B^2 \,.
}

\item The background determined by the metric \eqref{metric_17} in the basis 
\eqref{ex1_basis}
carries so-called geometric flux $f$,
which can be related to the structure constants $f_{ij}{}^k$ in the 
algebra of vector fields $\{e_i\}$, that is 
\eq{
  [ e_i, e_j] = f_{ij}{}^k e_k \,.
}
In the case of vanishing torsion,  these structure constants are proportional
to the coefficients of the curvature one-form, and in the present case they read
\eq{
  f_{23}{}^1 = - k^1\op h \,.
}

\item Finally, we note that the Killing vector \eqref{ex1_killing} corresponds to a compact isometry
of the three-torus only if the ratios $k^1:k^2:k^3$ are rational. If these are irrational, 
the isometry direction is non-compact. We exclude this  case here, 
although it might be interesting to study such configurations further.

\end{itemize}


\subsection{Twisted torus}
\label{sec_twisted_torus}

Let us now apply a T-duality transformation to the background obtained in the last section.
However, we do not start from 
the most general metric shown in \eqref{metric_17}, but only consider a circle bundle with a
rectangular two-torus as a base. The topology of this configuration is characterized by
\eq{
  \label{ex2_basis_19}
  d \mathsf e^1 = \mathsf f\op \mathsf e^2\wedge \mathsf e^3 \,,\hspace{60pt}
  d \mathsf e^2 = 0 \,, \hspace{60pt}
  d \mathsf e^3 = 0 \,,
}
where $\{\mathsf e^1,\mathsf e^2,\mathsf e^3\}$ is a globally-defined basis, and 
where $\mathsf f$ has dimensions
$\ell_{\rm s}^{-1}$. The geometry under consideration is specified by the following 
 metric
\eq{
  \label{ex2_metric_01}
  G_{ij} = \left( \begin{array}{ccc} 
  r_1^2 & 0 & 0 \\ 
  0 & r_2^2 & 0 \\
  0 & 0 & r_3^2
  \end{array} \right),
}  
and we allow for a non-vanishing field strength of the Kalb-Ramond field
\eq{
  \label{ex2_h-field}
  H = \mathsf h \, \mathsf e^1\wedge \mathsf e^2\wedge \mathsf e^3 \,,
  \hspace{80pt}
  \mathsf h \in \ell_{\rm s}^{-1}\op \mathbb Z \,.
}  
In order to apply the T-duality rules summarized in section~\ref{buscher_rules}, we have to go to 
a local frame, which we choose as
\eq{
  \label{ex2_coords}
  \mathsf e^{1} &= d Z^1 + \mathsf f\op\Bigl( \sigma_1 \op Z^2 dZ^3 - \sigma_2 \op Z^3 dZ^2 \Bigr)\,,
  \hspace{60pt} \sigma_1+\sigma_2=1 \,, \\
  \mathsf e^2 &= dZ^2 \,, \\[4pt]
  \mathsf e^3 &=dZ^3 \,.
}
Writing the metric \eqref{ex2_metric_01} in this local basis leads to a matrix $\tilde G_{ij}$ which is 
of a form similar to \eqref{ex1_matrix14}, and  which  we do not display here.
The topology  is specified by the following identifications in the coordinates $\{Z^i\}$
\eq{
  \arraycolsep2pt
  \renewcommand{\arraystretch}{1.3}
  \begin{array}{c@{\hspace{15pt}}lcl@{\hspace{20pt}}lcl}
  1) & Z^1 &\rightarrow&Z^1 +\ell_{\rm s} \,, \\
  2) & Z^2 &\rightarrow&Z^2 +\ell_{\rm s} \,, & Z^1 &\rightarrow&Z^1 - \ell_{\rm s}\op \mathsf  f\op\sigma_1\op Z^3 \,, \\
  3) & Z^3 &\rightarrow&Z^3 +\ell_{\rm s} \,, & Z^1 &\rightarrow&Z^1 + \ell_{\rm s}\op \mathsf f\op\sigma_2\op Z^2 \,.
  \end{array}
}


\subsubsection*{Killing vectors}

For the metric \eqref{ex2_metric_01}, there are six (local) Killing vectors describing  rotations and 
translations. Here, we are only interested in the isometries without fixed points,
which in the local coordinate system are given by
\eq{
  \label{ex2_killing}
  \arraycolsep2pt
  \tilde k_{(1)} = \left( \begin{array}{c} 1 \\ 0 \\ 0 \end{array} \right)  ,\hspace{40pt}
  \tilde k_{(2)} = \left( \begin{array}{c} -\sigma_1 \mathsf f \op Z^3 \\ 1 \\ 0 \end{array} \right)  ,\hspace{40pt}
  \tilde k_{(3)} = \left( \begin{array}{c} +\sigma_2 \mathsf f \op Z^2 \\ 0 \\ 1 \end{array} \right)  .
}  
Note that these vectors satisfy an algebra where the only non-vanishing commutator reads
\eq{
  \label{ex2_alg_kv_76}
  \bigl[ \tilde k_{(2)},\tilde k_{(3)} \bigr]_{\rm L} = \mathsf f \, \tilde k_{(1)} \,.
}
Now, suppose we want to perform a T-duality transformation along one of these directions.  
The vectors $\tilde k_{(a)}$ then have to satisfy the constraints summarize in
\eqref{restrictions} which, after an appropriate relabeling, read
\eq{
  \label{ex2_res}
  \begin{array}{l@{\hspace{40pt}}l}
  \tilde k_{(1)}: & \mbox{no restrictions} \,, \\[5pt]
  \tilde k_{(2)}: & \sigma_1 \op \mathsf f = 0 \,, \\[5pt]
  \tilde k_{(3)}: & \sigma_2 \op \mathsf f = 0 \,.
  \end{array}
}  
Therefore, applying our formalism and performing a T-duality transformation along say $\tilde k_{(2)}$
is only possible for geometries specified by $\sigma_1=0$ and $\sigma_2=1$;
 the opposite holds for $\tilde k_{(3)}$.

\pagebreak[2] 
Next,  to investigate the global properties of the vectors \eqref{ex2_killing},
let us switch from the local basis  to a basis of globally-defined vector fields.
To do so, we determine the following relations for the vector fields $\{\mathsf e_i\}$ dual to the one-forms 
$\{\mathsf e^i\}$
\eq{
  \partial_1 = \mathsf e_1 \,, \hspace{50pt}
  \partial_2 = \mathsf e_2 -  \sigma_2 \op \mathsf f\op Z^3 \op \mathsf e_1 \,, \hspace{50pt}
  \partial_2 = \mathsf e_3 + \sigma_1 \op \mathsf f \op Z^2 \op \mathsf e_1 \,.
}
This allows us to express the Killing vectors \eqref{ex2_killing} in a globally-defined basis as 
\eq{
  \label{ex2_kv}
  k_{(1)} & = \mathsf e_1 \,, \\
  k_{(2)} & = \mathsf e_2 - \mathsf f \op Z^3 \op \mathsf e_1 \,, \\
  k_{(3)} & = \mathsf e_3 + \mathsf f \op Z^2 \op \mathsf e_1 \,.
}
Note first that the choice of a local geometry has dropped out, that is, these vectors do not 
depend on $\sigma_1$ or $\sigma_2$. Second, and more importantly, we see
that $k_{(2)}$ and $k_{(3)}$ are not globally defined.
The quantity measuring the monodromy when going around the circle in the $Z^2$ or $Z^3$ direction 
is given by the geometric flux $\mathsf f$. In particular, we find
\eq{
  \label{ex2_kv_mono}
  &k_{(2)} \;\; \xrightarrow{\;\;Z^3\to Z^3+\ell_{\rm s}\;\;} \;\; k_{(2)} -\ell_{\rm s}\op\mathsf f \,  k_{(1)} \,, \\[3pt]
  &k_{(3)} \;\; \xrightarrow{\;\;Z^2\to Z^2+\ell_{\rm s}\;\;} \;\; k_{(3)} + \ell_{\rm s}\op\mathsf f \, k_{(1)} \,.
}
This implies that, strictly speaking, we are not allowed to perform a T-duality transformation along these
directions because, as one can infer for instance from \eqref{metric_10} and \eqref{field_strength}, the dual
metric and field strength will not be properly defined.


\subsubsection*{T-duality along $k_{(1)}$}

Since the Killing vector  $k_{(1)}$ is in fact globally defined,
we can perform a duality transformation along this direction unambiguously. Applying the transformation rules summarized in section~\ref{buscher_rules}, we first note that the dual basis of one-forms satisfies
\eq{
  \label{ex2_hm_00}
  d e^1 = \mathsf h\op e^2\wedge e^3 \,,\hspace{60pt}
  d e^2 = 0 \,, \hspace{60pt}
  d e^3 = 0 \,.
}
In this basis, the dual metric and the dual field strength read
\eq{
  \label{ex2_hm_01}
  \mathcal G_{ij} = \left( \begin{array}{ccc} 
  \frac{1}{r_1^2} & 0 & 0 \\ 
  0 & r_2^2 & 0 \\
  0 & 0 & r_3^2
  \end{array} \right), \hspace{80pt}
  \mathcal H= \mathsf f \, e^1\wedge e^2\wedge e^3 \,.
}  
Comparing with the initial configuration specified by 
\eqref{ex2_basis_19}, \eqref{ex2_metric_01} and \eqref{ex2_h-field}, 
we  therefore have, in agreement with \cite{Bouwknegt:2003vb,Bouwknegt:2003wp,Bouwknegt:2003zg},
that under T-duality 
\eq{
  \label{ex2_exchange}
  \mathsf f \;\longleftrightarrow \;\mathsf h \,, \hspace{80pt}
  r_1  \;\longrightarrow \; \frac{1}{r_1} \,.
}
Note furthermore, since $\mathcal H$ has to satisfy the quantization condition \eqref{quantization},  also the geometric flux $\mathsf f$ in \eqref{ex2_basis_19} has to be quantized as $\mathsf f\in\ell_{\rm s}^{-1} \mathbb Z$.


\subsubsection*{T-duality along $k_{(2)}$}

A more interesting situation is obtained by performing a 
T-duality transformation along the Killing vector $k_{(2)}$. As mentioned above,
this vector is not single valued on the twisted torus and hence 
T-duality is not well-defined. However, it has been argued in the literature \cite{Hellerman:2002ax} that one 
can nevertheless apply the transformation rules 
to obtain a locally-geometric background.

Our starting configuration is again specified by equations \eqref{ex2_basis_19}, \eqref{ex2_metric_01}
and \eqref{ex2_h-field}. However, due to the restrictions  \eqref{ex2_res},
we see that the only  local geometry which allows for a T-duality transform is 
determined by $\sigma_1=0$ and  
$\sigma_2=1$. In the  basis $\{dZ^1,dZ^2,dZ^3\}$ the initial metric is then expressed as follows
\eq{
  \renewcommand{\arraystretch}{1.3}
  \widetilde{G}_{ij} = \left( \begin{array}{ccc} 
  r_1^2 & -r_1^2 \op\mathsf f\op Z^3 & 0 \\ 
  -r_1^2 \op\mathsf f\op Z^3 & r_2^2 + r_1^2 (\op\mathsf f\op Z^3)^2 & 0 \\
  0 & 0 & r_3^2
  \end{array} \right).
}
After applying the duality along the direction  $\tilde k_{(2)} = (0,1,0)^T$,
the new basis of one-forms $\{E^1,E^2,E^3\}$ is characterized by
\eq{
  \label{ex2_basis_29}
  d E^1 = 0 \,, \hspace{60pt}  
  d E^2 = - \mathsf h\op E^1\wedge E^3 \,,\hspace{60pt}
  d E^3 = 0 \,.
}
The T-dual metric and field strength in this basis read
\eq{
  \label{ex2_gh}
  \renewcommand{\arraystretch}{1.3}
  &\mathcal G_{ij} = 
  \left( \begin{array}{ccc} 
   r_1^2 \,\eta & 0 & 0 \\ 
  0 &  \frac{1}{r_2^2}\, \eta  & 0 \\
  0 & 0 & r_3^2
  \end{array} \right),
  \hspace{110pt}
  \eta = \frac{1}{1+\bigl(\frac{r_1}{r_2} \op\mathsf f Z^3\bigr)^2} \,, \\[5pt]
  &\mathcal H = - \mathsf f\left( \frac{r_1}{r_2}\right)^2
  \eta^2 \left[ 1-\bigl(\tfrac{r_1}{r_2}\op \mathsf f Z^3\bigr)^2 \right]  E^1\wedge E^2\wedge E^3\,.
}  
Here we see that because the Killing vector $k_{(2)}$ was not globally defined, also the dual 
metric and field strength are  valid only locally.
However, as shown in \cite{Dabholkar:2002sy,Hull:2004in}, this space can be 
promoted to a  global background if in addition to diffeomorphisms one considers T-duality transformations 
as transition functions.


\subsubsection*{Transition functions}

The background shown in \eqref{ex2_gh} is  called a T-fold \cite{Hull:2004in}, 
and is conveniently described in terms of a doubled geometry. 
We do not want to review this discussion here, but  present
an interpretation from a slightly different point of view.

As we can see from the formulas in equation \eqref{ex2_gh}, when going around the circle in the 
$Z^3$-direction, the metric and field strength are not periodic. Note that the mismatch cannot be compensated 
by using local coordinate charts and applying diffeomorphisms as transition functions.
Similarly, since the metric and field strength are expected to be gauge invariant quantities, 
gauge transformation do not play a role either.
However, as noted in \cite{Dabholkar:2002sy,Hull:2004in}, a third possibility is to employ 
T-duality transformations as transition functions.
Let us therefore recall from \eqref{ex2_exchange} that 
a single T-duality typically interchanges the $H$-flux with the topological twisting $\mathsf f$, and that it inverts the 
radius.
And indeed,  a T-duality transformation along the $e^1$-direction for \eqref{ex2_gh} 
results in a background given by substituting  
\eq{
  \mathsf f \;\longleftrightarrow \;\mathsf h \,, \hspace{80pt}
  r_1  \;\longrightarrow \; \frac{1}{r_1} \,.
}
Thus, in order to make the T-fold  globally defined, an even number of T-duality transformations
has to be performed.
\begin{figure}[t]
\centering
\begin{picture}(0,0)
\put(-3,-8){\footnotesize$\Phi_{AB}$}
\put(-5,-160){\footnotesize$\phi_{12}$}
\put(-78,-138){\footnotesize$\mathsf T$}
\put(58,-138){\footnotesize$\mathsf T$}
\put(90,-235){\scriptsize twisted torus}
\put(135,-25){\scriptsize T-fold}
\put(-45,-48){\scriptsize$A$}
\put(40,-48){\scriptsize$B$}
\put(-49,-215){\scriptsize$1$}
\put(38,-215){\scriptsize$2$}
\end{picture} \\
\includegraphics[width=0.65\textwidth]{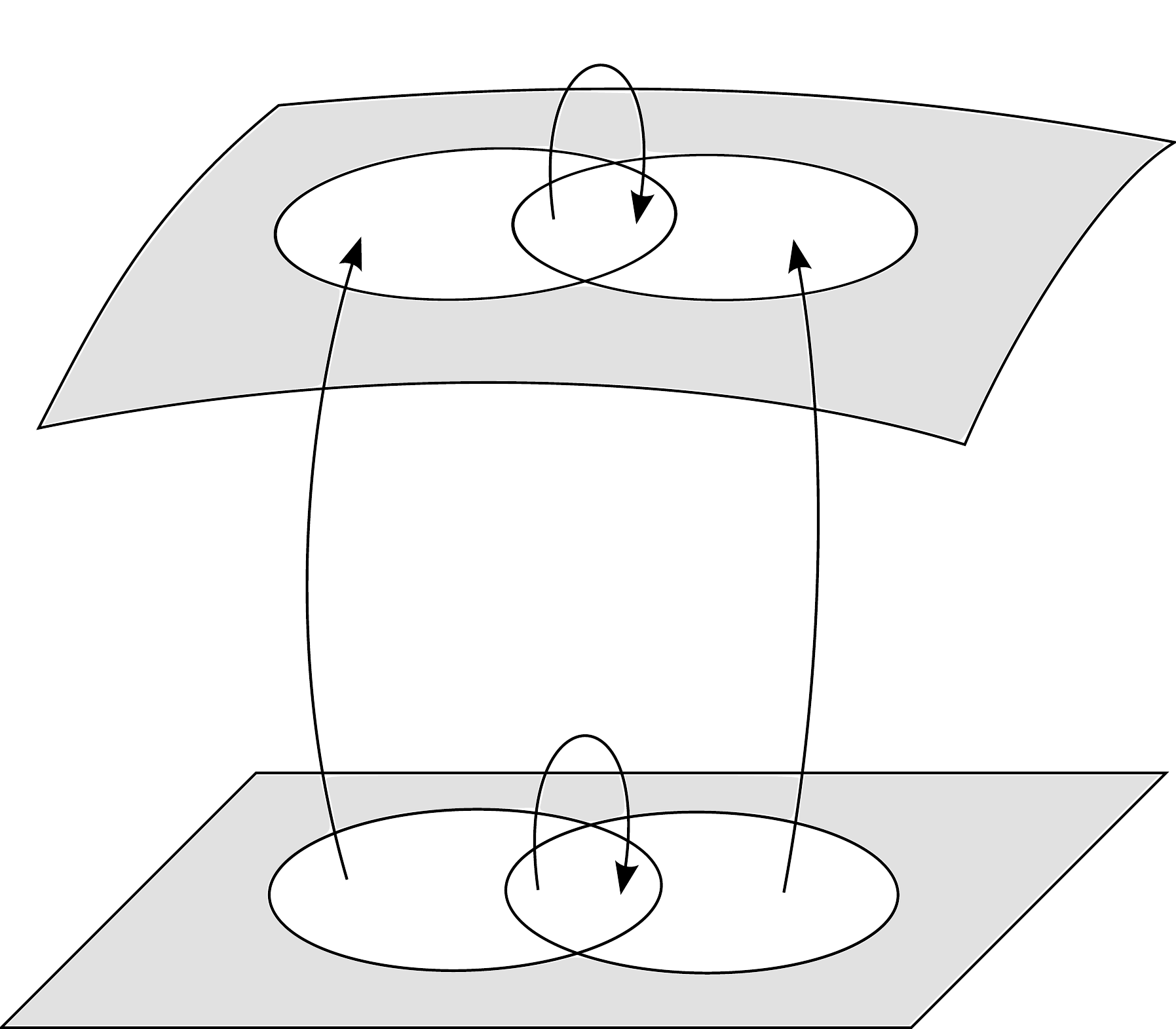} \\
\vspace{5pt}
\caption{Illustration of a T-fold. The transition function $\Phi_{AB}$ between two local charts
on the T-fold can be interpreted as the 
composition $\Phi_{AB} = \mathsf T\circ\phi_{12}\circ\mathsf T^{-1}$, where $\mathsf T$ denotes
a T-duality transformation and where $\phi_{12}$ is the
(geometric) transition function on the twisted torus.\label{fig_05}}
\end{figure}
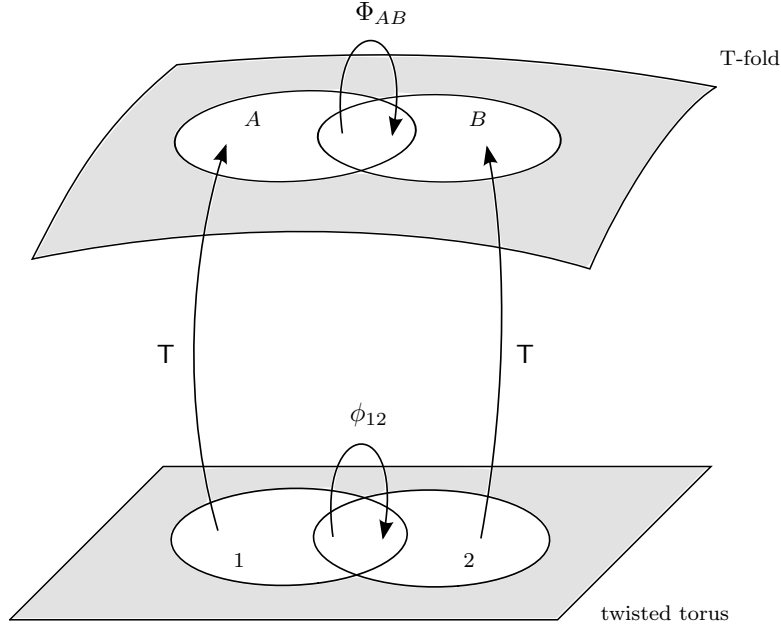
This suggests that a transition function $\Phi_{AB}$ between local overlapping charts 
on a  T-fold should be given by a combination of 
a T-duality back to the twisted torus $\mathsf T^{-1}$, a diffeomorphism $\phi_{12}$ on the twisted torus, and a second T-duality $\mathsf T$ to the T-fold. This is illustrated in figure~\ref{fig_05}, which in formulas 
reads  as follows
\eq{
  \Phi_{AB} = \mathsf T_{k_{(2)}}\circ\phi_{12}\circ\mathsf T^{-1}_{k_{(2)}-\mathsf f \op\ell_{\rm s}\op k_{(1)}}\,.
}
Note that here it is important that the monodromy of the Killing vector is again a Killing vector, which allows us to 
perform the first T-duality transformation back to the twisted torus.


\subsection{T-fold}
\label{sec_tfold}

We finally want to briefly remark on the isometry structure of the T-fold. 
Employing a  basis of vector fields $\{E_i\}$ dual to \eqref{ex2_basis_29},
the (local) Killing vectors corresponding
to the metric in \eqref{ex2_gh} are determined as
\eq{
  \label{ex3_killing}
  \arraycolsep2pt
  K_{(1)} = E_1 + \mathsf h\op Z^3 E_2  ,\hspace{80pt}
  K_{(2)} = E_2\, .
}  
In contrast
to the twisted torus with the three linearly independent  vectors shown in \eqref{ex2_kv},
 the T-fold admits only two Killing vectors.
This is sometimes mentioned as a puzzle, however, it is easily understood recalling our discussion
on page \pageref{mult_glob}. 
In particular, to arrive at the T-fold background, we have performed
a duality transformation along $k_{(2)}$
on the twisted torus defined in the beginning of section~\ref{sec_twisted_torus}. 
The Killing vectors of the latter satisfy  the algebra
\eqref{ex2_alg_kv_76}, which we recall for convenience
\eq{
  \bigl[ k_{(1)}, k_{(2)} \bigr]_{\rm L} = 0 \,, \hspace{40pt}
  \bigl[ k_{(2)}, k_{(3)} \bigr]_{\rm L} = \mathsf f \,  k_{(1)} \,,  \hspace{40pt}
  \bigl[ k_{(3)}, k_{(1)} \bigr]_{\rm L} = 0 \,.  
}
As explained on page \pageref{mult_glob}, 
when gauging the sigma-model action, the surviving global symmetries are 
determined by \eqref{constr_02}. Since $k_{(2)}$  commutes with $k_{(1)}$ but not with $k_{(3)}$,
the dual geometry is not expected to have $k_{(3)}$ as an isometry. And indeed, the T-fold 
possesses only the Killing vectors shown in \eqref{ex3_killing}.


\subsubsection*{Remark}

As we mentioned earlier, it has been argued \cite{Shelton:2005cf} that formally an additional T-duality
transformation on the T-fold along the $E_3$-direction can be performed, even though this 
does not correspond to a Killing vector. The resulting space is called an
$R$-flux background, which has a non-associative structure \cite{Bouwknegt:2004ap,Bouwknegt:2004tr,Ellwood:2006my}. This T-duality procedure is best discussed in the context
of a doubled geometry, which we are not going to review here.
However, it would be interesting to further investigate this situation from a sigma-model point of view.


\vspace*{1.5em}

\pagebreak

\section{Examples II -- Spheres}
\label{sec_spheres}

After having discussed in detail T-duality transformations for the torus, we now 
turn to  the sphere. First, we review and extend the 
work in \cite{Alvarez:1993qi} about abelian T-duality for the three-sphere, 
and then present an explicit example of a non-geometric T-fold in section~\ref{sec_tw_sphere}.


\subsection{Sphere}
\label{sec_sphere}

We start by specifying the geometry of the three-sphere. A coordinate 
system  convenient for our purposes is that of Hopf coordinates, describing
the embedding of $S^3$ into $\mathbb C^2$.
More concretely, let us chose the following complex coordinates
\eq{
  \label{hopf_01}
  z_1 &=  R\, e^{i\op\xi_1} \sin\eta\,, \\
  z_2 &=  R\, e^{i\op\xi_2} \cos\eta\,,
}
with $\xi_{1,2}= 0\ldots 2\pi$ and $\eta=0\ldots \pi/2$, where $R$ denotes the radius of the three-sphere.\footnote{Note that in order to simplify the formulas in this section, we have set $\ell_{\rm s}=1$. The dimensions can  be re-installed by assigning $\ell_{\rm s}$ to $\xi_{1}$, $\xi_2$ and $\eta$.}
The latter is then described by the constraint
$R^2 = |z_1|^2 + |z_2|^2$,
and the round metric in the coordinates \eqref{hopf_01} is characterized by the following line-element squared
\eq{
  \label{hopf_02}
  ds^2 = R^2 \,\Bigl[ \,\sin^2\eta \,(d\xi_1)^2
  + \cos^2\eta \,(d\xi_2)^2
  + (d\eta)^2\, \Bigr]\,.
}
Note that  the metric becomes singular at $\eta=0$ and $\eta=\pi/2$, and hence one should 
use different coordinate charts in the neighborhoods of these points. However, for simplicity 
here we  work with only one coordinate patch, 
but we are careful in avoiding the singular points in our analysis.
Furthermore, we also consider a non-trivial field strength 
\eq{
  H = \frac{h}{2\pi^2}\op \sin \eta \cos\eta \, d\xi_1\wedge d\xi_2 \wedge d\eta 
}
for the Kalb-Ramond field $B$,
for which the quantization condition in equation \eqref{quantization} implies that
$h \in  \mathbb Z$.


\subsubsection*{Killing vectors}

Next, we turn  to the isometries  of the three-sphere. 
The isometry group for $S^3$ is $O(4)$, and thus there are six linearly independent Killing vectors.
Employing the basis of vector fields $\{\partial_{\xi_1},\partial_{\xi_2},\partial_{\eta}\}$, the Killing vectors can
be expressed in the following way
\eq{
  \label{hopf_killing}
  \arraycolsep2pt
  \begin{array}{ll@{\hspace{40pt}}ll}
  k_{(1)} &\displaystyle =\left(\begin{array}{c} 1 \\ 0 \\ 0 \end{array}\right), &
  k_{(2)} &\displaystyle =\left(\begin{array}{c} 0 \\ 1 \\ 0 \end{array}\right), 
  \\[22.5pt]  
  k_{(3)} &\displaystyle =\left(\begin{array}{c} -\sin\xi_1 \cos\xi_2 \cot \eta  \\ +\cos\xi_1 \sin\xi_2 \tan\eta 
     \\ \cos\xi_1\cos\xi_2 \end{array}\right), &
  k_{(4)} &\displaystyle =\left(\begin{array}{c} -\sin\xi_1 \sin\xi_2 \cot \eta  \\ -\cos\xi_1 \cos\xi_2 \tan\eta 
     \\ \cos\xi_1\sin\xi_2 \end{array}\right), 
  \\[22.5pt]  
  k_{(5)} &\displaystyle =\left(\begin{array}{c} +\cos\xi_1 \cos\xi_2 \cot \eta  \\ +\sin\xi_1 \sin\xi_2 \tan\eta 
     \\ \sin\xi_1\cos\xi_2 \end{array}\right), &
  k_{(6)} &\displaystyle =\left(\begin{array}{c} +\cos\xi_1 \sin\xi_2 \cot \eta  \\ -\sin\xi_1 \cos\xi_2 \tan\eta 
     \\ \sin\xi_1\sin\xi_2 \end{array}\right).
  \end{array}
}
As one can check, these indeed satisfy the commutation relations of the algebra
$\mathfrak{so}(4)$. 
Furthermore, let us observe that out of the vectors in \eqref{hopf_killing} we can form three linear combinations 
\eq{
  \label{hopf_03}
  e_{(1)} = k_{(1)} - k_{(2)} \,,\hspace{40pt}
  e_{(2)} = k_{(3)} + k_{(6)} \,,\hspace{40pt}
  e_{(3)} = k_{(4)} - k_{(5)} \,,
}
which are orthogonal to each other with respect to the metric \eqref{hopf_02}, and which satisfy
\eq{
  |e_{(a)}|^2 = R^2 \,, \hspace{80pt}
  [ e_{(a)} , e_{(b)} ]_{\rm L} = 2\op \epsilon_{ab}{}^c \op e_{(c)} \,,
}
for $a,b,c\in\{1,2,3\}$ and  $\epsilon_{abc}$ the Levi-Civita symbol.
In particular, the vector fields \eqref{hopf_03} are nowhere vanishing, 
corresponding to the fact that they  are 
dual to the left-invariant one-forms on the three-sphere.


\subsubsection*{T-duality}

As we mentioned above, at $\eta=0$ and $\eta=\pi/2$ the metric \eqref{hopf_02} becomes singular.
When performing a T-duality transformation on $S^3$,  we therefore consider only linear combinations of 
$k_{(1)}$ and $k_{(2)}$, which are independent of $\eta$. More concretely, let us write
\eq{
  \label{hopf_kv_01}
  k = \alpha\op k_{(1)} + \beta \op k_{(2)} = \alpha\op \partial_{\xi_1} + \beta\op \partial_{\xi_2} 
}
for $\alpha\neq0$ and $\alpha,\beta={\rm const}$.
This Killing vector satisfies $\mathcal L_k H = 0$ as well as
the constraints shown in \eqref{restrictions}.
Recalling then our discussion on the T-duality transformation rules from page \pageref{buscher_page}, we see that
the dual metric and field strength
are expressed in a basis $\{E^{\chi},E^{\xi},E^{\eta}\}$ of one-forms characterized by
\eq{
  \label{hopf_basis_11}
  dE^{\chi} = \frac{\alpha \op h}{2\pi^2}\, \sin\eta\cos\eta \,d\xi\wedge d\eta \,,
  \hspace{40pt}
  dE^{\xi} = 0\,, \hspace{40pt}
  dE^{\eta} =0 \,.
}
Note that here we simplified our notation by replacing $\xi_2\to \xi$.
The norm of $k$ is given by $|k|^2 = R^2(\alpha^2 \sin^2\eta + \beta^2\cos^2 \eta)$,
and the components of the metric in the above basis read
\eq{
  \label{hopf_metric_19}
  \mathcal G_{ij} =  \left( \begin{array}{ccc} 
  \frac{1}{R^2}  \rho& 0 & 0 \\ 
  0 & \frac{1}{4}R^2 \rho\op\alpha^2 \sin^2(2\eta)& 0 \\
  0 & 0 & R^2
  \end{array} \right),
  \hspace{35pt}
  \rho=\frac{1}{\alpha^2\op \sin^2 \eta + \beta^2 \cos^2\eta}\,.
}
For the dual field strength of the Kalb-Ramond field, we deduce from equation \eqref{buscher_02} that
\eq{
  \label{ex3_h_94}
  \mathcal H = \alpha^2\beta \,\frac{\sin(2\eta)}{(\alpha^2 \cos^2\eta + \beta^2 \sin^2\eta)^2}
  \,E^{\chi}\wedge E^{\xi}\wedge E^{\eta}
  \,.
}  
The topology of this space can be understood by 
recalling from equation \eqref{math_02} that the first Chern class of the dual background is
determined by the $H$-flux of the initial configuration. This implies that for $h\neq 0$, we obtain a non-trivial circle bundle
over $S^2$ (for more details see for instance section 4.3 in \cite{Bouwknegt:2003vb}).
Now, in order to better understand the geometry of the above space, let us introduce a local basis which
satisfies \eqref{hopf_basis_11}. We write
\eq{
  \label{ex3_metric_395}
  E^{\chi} &= d\chi + \frac{\alpha\op h}{2\pi^2} \Bigl( \op\tau_1\xi \sin\eta\cos\eta \op d\eta
  + \tfrac{1}{2}\op\tau_2 \cos^2\eta \, d\xi \Bigr) \,, \hspace{40pt}\tau_1+\tau_2 =1 \,,
  \\
  E^{\xi} & = d\xi \,, \\[5pt]
  E^{\eta} & = d\eta \,,
}
with $\tau_1$ and $\tau_2$  constant.
This results in a rather complicated background which is essentially specified by the  parameters 
$\alpha/\beta$, $h$ and $\tau_1=1-\tau_2$. 
We do not want to discuss this space in full generality, but only focus
on some specific cases in the 
following.


\subsubsection*{Backgrounds with $\alpha^2=\beta^2$}

Let us begin with  $\alpha=\beta=1$ and  $h=0$. In this situation the T-dual background
corresponds to $S^2\times S^1$,
as it has already been discussed for instance in  \cite{Alvarez:1993qi}.
Indeed, after rescaling $\eta\to \eta/2$, we see 
from \eqref{hopf_metric_19}
that the line-element squared reads
\eq{
  \mathsf{ds}^2 = \frac{R^2 }{4}\, \Bigl[ \,(d\eta)^2 +\sin^2\eta\, (d\xi)^2\, \Bigr] + \frac{1}{R^2}(d\chi)^2 \,,
}
for $\eta=0\ldots \pi$ and $\xi=0\ldots 2\pi$. The dual field strength, after rescaling $\eta$,
becomes
\eq{
  \mathcal H = \frac{1}{2}\sin\eta\,d\eta \wedge d\xi\wedge d\chi \,.
}
For $h\neq 0$, the  geometry is that of a circle in the $\chi$-direction which is non-trivially
fibered over the two-sphere. The corresponding 
metric in local coordinates can be determined from \eqref{ex3_metric_395},
but we will not present the explicit expression here.

More evidence that the T-dual geometry indeed corresponds to a circle bundle over $S^2$ 
(see also \cite{Bouwknegt:2003vb}) is 
provided by the structure of the isometry group. 
Let us therefore recall  our discussion from page \pageref{mult_glob}
about the remaining global symmetries after a T-duality transformation. 
They  are described by those Killing vectors
in \eqref{hopf_killing} which commute with \eqref{hopf_kv_01},
and for $\alpha=\beta=1$ they are given by \eqref{hopf_kv_01} and \eqref{hopf_03}, that is
\eq{
  \label{ex3_kv_after_83}
  \arraycolsep1.8pt
  \begin{array}{ll@{\hspace{50pt}}ll}
   e_{(0)} &=  k_{(1)} + k_{(2)} \,, & 
   e_{(1)} &=  k_{(1)} - k_{(2)} \,, \\[4pt]
   e_{(2)} &=  k_{(3)} + k_{(6)} \,,&
   e_{(3)} &=  k_{(4)} - k_{(5)} \,.
  \end{array}
}
These satisfy, as expected,  the $\mathfrak{so}(3)\times \mathfrak u(1)$ isometry algebra of $S^2\times S^1$.
In particular,  we find the commutation relations
\eq{
  [ e_{(0)},  e_{(a)} ]_{\rm L} = 0\,,\hspace{50pt}
  [ e_{(a)},  e_{(b)} ]_{\rm L} = 2 \op\epsilon_{ab}{}^c \op e_{(c)}\,,
}
for $a,b,c\in\{1,2,3\}$.
Hence, the vectors \eqref{ex3_kv_after_83}  correspond to symmetries of the dual sigma-model.
This can be verified also explicitly by computing the Killing vectors for the T-dual background. In the
basis  $\{E_{\chi},E_{\xi},E_{\eta}\}$ dual to the one-forms characterized by \eqref{hopf_basis_11},
the directions of isometry of \eqref{hopf_metric_19} (for $\alpha=\beta=1$) are given by
\eq{
  \label{ex_3_hvm}
  K_{(0)} = E_{\chi} \,, \hspace{40pt}
  K_{(1)} &=E_{\xi} + \frac{h}{8\pi^2}\cos(2\eta) E_{\chi}\,, \\
  K_{(2)} &=+\frac{1}{2}\cos\xi \,E_{\eta}-\frac{\sin \xi}{\tan(2\eta)}\op E_{\xi} + \frac{h}{8\pi^2} 
     \sin\xi\sin(2\eta) E_{\chi} \,, \\
  K_{(3)} &=-\frac{1}{2}\sin\xi \,E_{\eta}-\frac{\cos \xi}{\tan(2\eta)}\op E_{\xi} + \frac{h}{8\pi^2} 
     \cos\xi\sin(2\eta) E_{\chi} \,.
}
These  vectors satisfy $\mathcal L_{K_{(0)}}\mathcal G=\mathcal L_{K_{(a)}}\mathcal G=0$ as well as 
$\mathcal L_{K_{(0)}}\mathcal H=\mathcal L_{K_{(a)}}\mathcal H=0$,
and generate  the $\mathfrak{so}(3)\times \mathfrak u(1)$ isometry algebra
\eq{
  [ K_{(0)},  K_{(a)} ]_{\rm L} = 0\,,\hspace{50pt}
  [ K_{(a)},  K_{(b)} ]_{\rm L} = \epsilon_{ab}{}^c \op K_{(c)}\,.
}
We therefore have  a second non-trivial check of our result 
on page \pageref{mult_glob} about the remaining global 
symmetries after a T-duality transformation.


\subsubsection*{Backgrounds with $\alpha^2\neq\beta^2$}

\begin{figure}[p]
\centering
\subfigcapskip15pt
\begin{tabular}{@{}l@{\hspace{55pt}}r@{}}
\subfigure[$\beta=5$]{
\includegraphics[height=180pt]{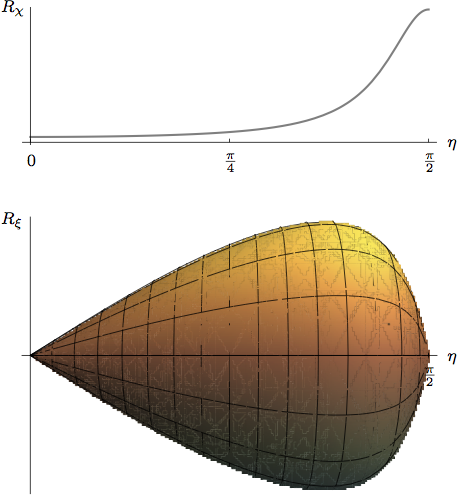}
\label{fig_05_5}
} 
&
\subfigure[$\beta=4/5$]{
\includegraphics[height=180pt]{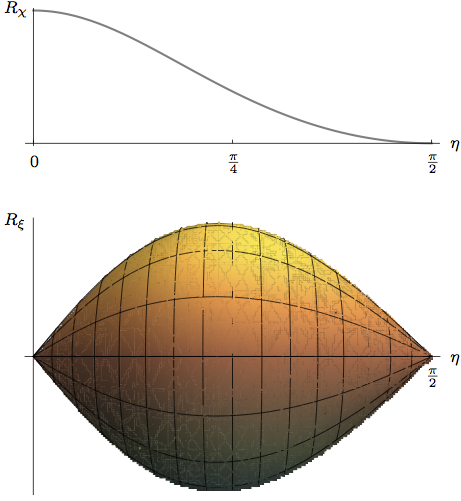}
\label{fig_05_45}
} 
\\[65pt]
\subfigure[$\beta=1/5$]{
\includegraphics[height=180pt]{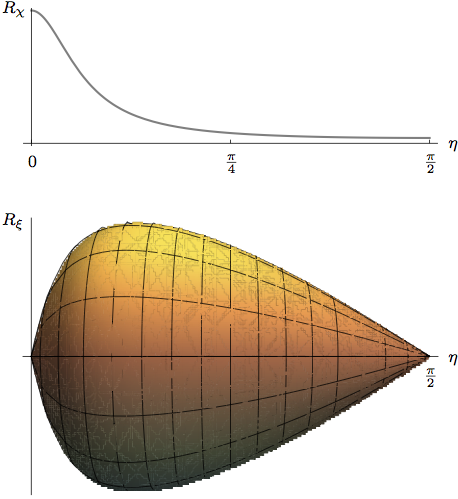}
\label{fig_05_15}
} 
&
\subfigure[$\beta=0$]{
\includegraphics[height=180pt]{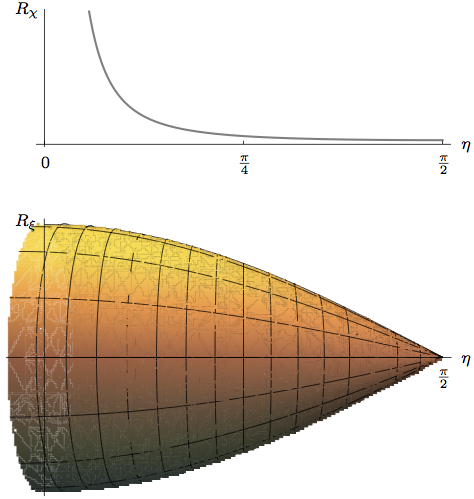}
\label{fig_05_0}
} 
\end{tabular}
\caption{Illustration of the space described by the metric \eqref{hopf_76} for $R=1$ and 
different values of $\beta$. The lower three-dimensional plot shows the space 
parametrized by $\eta$ and $\xi$, and the
upper plot shows the dependence  of the radius for $\chi$ on $\eta$. 
Note that in (d) the value of $R_{\xi}$ is finite at $\eta=0$ but $R_{\chi}$ diverges.}
\end{figure}

The next case we want to consider is $h=0$, $\alpha=1$ and $\beta^2\neq\alpha^2$. For $\beta$ non-vanishing,
the dual metric is non-singular and the corresponding line-element squared reads
\eq{
  \label{hopf_76}
  \mathsf{ds}^2 = R^2(d\eta)^2
  +\frac{1 }{\sin^2\eta + \beta^2 \cos^2\eta} \left[ \frac{1}{R^2} (d\chi)^2 + \frac{R^2}{4} \sin^2(2\eta)
  (d\xi)^2
  \right].
}
This describes a circle fibered non-trivially over a deformed sphere, which is illustrated schematically in figures~\ref{fig_05_5}, \ref{fig_05_45} and \ref{fig_05_15}.
The dual field strength is given by
\eq{
  \mathcal H = \beta \,\frac{\sin(2\eta)}{(\cos^2\eta + \beta^2 \sin^2\eta)^2}
  \,d\chi\wedge d\xi\wedge d\eta
  \,.
}  
For $h=0$, $\alpha=1$ and $\beta=0$ the resulting space becomes singular since
the length of the Killing vector vanishes at $\eta=0$,
as illustrated in figure~\ref{fig_05_0}.
Finally, for $h\neq0$ and $\alpha^2\neq \beta^2$ the  geometry 
is rather complicated and we do not present a detailed analysis here.
Before closing this section, let us also comment on the isometries for the T-dual background and 
remark that for $\alpha^2\neq \beta^2$ 
only two of the Killing vectors  in \eqref{hopf_killing}  commute with \eqref{hopf_kv_01}, namely
\eq{
  \label{ex3_kv_948}
  k_{(1)} \hspace{40pt}\mbox{and}\hspace{40pt}k_{(2)} \,.
}
This agrees with the fact that the metric \eqref{hopf_76} has only two directions of isometry, described by $k_{(\chi)}=\partial_{\chi}$ and $k_{(\xi)}=\partial_{\xi}$.


\subsection{Twisted sphere}
\label{sec_tw_sphere}

Our aim in this section is to construct a non-geometric background via a T-duality 
transformation on the sphere. However, as we
have seen in the last section, for the three-sphere  the Killing vectors \eqref{hopf_killing}
are all globally defined. Similarly, after applying a T-duality transformation to $S^3$, also the Killing vectors of the
resulting background shown in \eqref{ex_3_hvm} and below \eqref{ex3_kv_948} do not exhibit monodromies of the form \eqref{ex2_kv_mono}. Hence, we do not expect to find a non-geometric background by applying successive 
T-duality transformations to a single three-sphere.\footnote{Following the reasoning of \cite{Alvarez:1993qi},
this implies that, at least for a single three-sphere $S^3$, no non-geometric fluxes should appear 
in the corresponding dimensionally-reduced theory. }

But, let us recall from equation \eqref{hopf_02} that the three-sphere can be interpreted as a two-torus fibered over
a line segment. If we then consider a four-dimensional space in which a circle is twisted over $S^3$, we
arrive at a situation similar to the twisted torus discussed in 
section~\ref{sec_twisted_torus}.\footnote{Note that, in contrast to the twisted torus, 
the second cohomology class of $S^3$ is trivial, and hence all circle bundles over $S^3$ are topological trivial.
Nevertheless, a non-geometric background can be obtained. We thank the referee at JHEP
for pointing this out to us.}
Therefore, in this case it appears to be possible to obtain a T-fold after an appropriate T-duality transformation.


\subsubsection*{Geometry}

Let us  specify the geometry of the four-dimensional space. Compared to section~\ref{sec_sphere},
we slightly change the parametrization of $S^3$ by substituting $\xi_{1,2}\to\xi_1\pm\xi_2$ 
 in \eqref{hopf_02}.
The metric under consideration is then given by the
 following line-element squared
\eq{
  \label{ex4_hvm}
  ds^2 = R_1^2 \,\Bigl[ \,\bigl(\mathsf E^{\xi_1}\bigr)^2
  -2 \cos(2\eta)\, \mathsf E^{\xi_1}\!\otimes \mathsf E^{\xi_2}
  + \bigl(\mathsf E^{\xi_2}\bigr)^2
  + (\mathsf E^{\eta})^2\, \Bigr]
  +R^2_2 \bigl( \mathsf E^{X} \bigr)^2 
  \,.
}
The topology of the space is characterized by the  one-forms $\{\mathsf E^{\xi_1},\mathsf E^{\xi_2},\mathsf E^{\eta},\mathsf E^{X} \}$ which satisfy the algebra 
\eq{
  d \op\mathsf E^{\xi_1} =
  d \op\mathsf E^{\xi_2} =
  d\op \mathsf E^{\eta} =0 \,, 
  \hspace{60pt}
   d\op \mathsf E^X = f\op \mathsf E^{\xi_1} \wedge \mathsf E^{\xi_2} \,.
}
In local coordinates, these one-forms can be expressed as
\eq{
  \label{ex3_blub}
  \arraycolsep2pt
  \renewcommand{\arraystretch}{1.3}
  \begin{array}{ll}
  \displaystyle \mathsf E^{\xi_1} & = d \xi_1 \,, \\
  \displaystyle \mathsf E^{\xi_2} & = d \xi_2 \,, \\
  \displaystyle \mathsf E^{\eta} & = d \eta \,, \\  
  \displaystyle \mathsf E^{X} & 
  \displaystyle = d X + f\bigl( \rho_1 \op\xi_1 \op d\xi_2 - \rho_2 \op \xi_2 \op d\xi_1\bigr)\,, 
  \hspace{50pt} \rho_1+\rho_2=1 \,,
  \end{array}
}
where the choice of local coordinates is parametrized by the constants $\rho_1$ and $\rho_2$.
Furthermore, we observe that if we require the one-forms \eqref{ex3_blub}
to be well-defined, we have to make the identifications
\eq{
  \label{ex3_ident}
  \arraycolsep2pt
  \renewcommand{\arraystretch}{1.3}
  \begin{array}{c@{\hspace{15pt}}lcl@{\hspace{30pt}}lcl}
  1) & \xi_1 &\rightarrow&\xi_1 +2\pi \,, & X &\rightarrow& X- 2\pi \op f\rho_1\op\xi_2 \,, \\
  2) & \xi_2 &\rightarrow&\xi_2 +2\pi \,, & X &\rightarrow& X+ 2\pi \op f\rho_2\op\xi_1 \,,\\
  3) & X &\rightarrow& X+ 2\pi \,.
  \end{array}
}
For our purposes in the following,  three Killing vectors of the metric \eqref{ex4_hvm} are of 
interest. In terms of the  vector fields $\{\mathsf E_{\xi_1},\mathsf E_{\xi_2},\mathsf E_{\eta},\mathsf E_{X} \}$, dual to the corresponding one-forms, we have
\eq{
  \label{ex4_kv_global}
  \mathsf K_{(1)} = \mathsf E_{X} \,, \hspace{60pt}
  \mathsf K_{(2)} &=\mathsf E_{\xi_1} - f\op\xi_2 \op \mathsf E_{X} \,, \\[3pt]
  \mathsf K_{(3)} &=\mathsf E_{\xi_2} + f\op\xi_1 \op \mathsf E_{X} \,.
}
Note that these Killing vectors are not all globally defined. Indeed, under the identifications \eqref{ex3_ident}
we find the following  monodromies
\eq{
  \label{ex3_monodd}
  &\mathsf K_{(2)} \;\; \xrightarrow{\;\;\xi_2\to \xi_2+2\pi\;\;} \;\; \mathsf K_{(2)} -2\pi\op f \,  \mathsf K_{(1)} \,, \\[3pt]
  &\mathsf K_{(3)} \;\; \xrightarrow{\;\;\xi_1\to \xi_1+2\pi\;\;} \;\; \mathsf K_{(3)} + 2\pi\op f \, \mathsf K_{(1)} \,.
}


\subsubsection*{T-duality along $\mathsf K_{(1)}$}

Let us now consider a T-duality transformation along $\mathsf K_{(1)}$ on the above background. 
The metric is determined by \eqref{ex4_hvm}, and we choose a vanishing field strength 
for the Kalb-Ramond field. 
In order to apply the T-duality rules from page \pageref{buscher_page}, we 
first have to check whether the restrictions \eqref{restrictions} are satisfied.
This can be done by expressing $\mathsf K_{(1)}$ in a local basis as follows
\eq{
  \mathsf K_{(1)} = \partial_X \,,
}
which indeed solves the above-mentioned constraints. When determining the dual metric, also the 
metric tensor has to be written in the  basis $\{d\xi_1,d\xi_2,d\eta,dX\}$,
resulting in a rather lengthy expression which we do not display here. However, employing 
\eqref{buscher_1} together with \eqref{buscher_03}, we find the following dual line-element squared
\eq{
  \mathsf{ds}^2 = R_1^2 \,\Bigl[ \,(d\xi_1)^2
  -2 \cos(2\eta) \,d\xi_1\otimes d\xi_2
  +(d\xi_2)^2
  + (d\eta)^2\, \Bigr]+ \frac{1}{R_2^2}\, (d\chi)^2\,.
}
The T-dual field strength can be determined from equation \eqref{buscher_02} and reads 
\eq{
  \mathcal H = f\op d\xi_1\wedge d\xi_2\wedge d\chi\,.
}  
Therefore, the background obtained after applying a T-duality transformation on the twisted sphere
along $\mathsf K_{(1)}$ is given by $S^3\times S^1$, with a  non-vanishing field strength $\mathcal H$.


\subsubsection*{T-duality along $\mathsf K_{(2)}$}

We now turn to the Killing vector  $\mathsf K_{(2)}$. Even though $\mathsf K_{(2)}$ is not single valued,
we proceed along similar lines as in section~\ref{sec_twisted_torus}
and  perform a T-duality transformation on the geometry \eqref{ex4_hvm} locally. This then leads to a
non-geometric space, which is only locally geometric.
Note that the field strength $H$ of the Kalb-Ramond field is again chosen to be vanishing.

As in the previous examples, in order to apply the transformation rules,
we have to express the metric and Killing vectors in local coordinates. For the latter,
we find
\eq{
  \mathsf K_{(2)} = \partial_{\xi_1} - f\rho_1\op\xi_2\op\partial_X\,,
}
and thus the constraints in \eqref{restrictions} are only satisfied for
geometries specified by $\rho_1=0$ and $\rho_2=1$. In this case, the metric in local 
coordinates $\{d\xi_1,d\xi_2,d\eta,dX\}$ simplifies and takes the following form
\eq{
  \renewcommand{\arraystretch}{1.2}
  \widetilde G_{ij} = \left( \begin{array}{cccc}
  R_1^2 + R_2^2 \op f^2 \xi_2^2 & -R_1^2 \cos(2\eta) & 0 & -R_2^2 f \xi_2 \\
  -R_1^2 \cos(2\eta) & R_1^2 & 0 & 0 \\
  0 & 0 & R_1^2 & 0 \\
  -R_2^2 f \xi_2 & 0  & 0 & R_2^2 
  \end{array}
  \right) ,
}
where $i,j\in\{\xi_1,\xi_2,\eta,X\}$.
Applying then the transformation rules given in equation \eqref{buscher_1}, we find the following dual
line-element squared 
\eq{
  \label{ex3_t-geom}
  \mathsf{ds}^2 =
  R_1^2 \Bigl[  \op(d\eta)^2 + \sin^2(2\eta) (d\xi)^2 \op\Bigr] +
  \vartheta(\xi)
  \left[\,
  \tfrac{1}{R_1^2}\,(d\chi)^2 + R_2^2 \op\Bigl( dX - \cos(2\eta) f\op\xi \op d\xi \Bigr)^2 \right] 
  ,
}
where we simplified our notation by replacing $\xi_2\to \xi$, and where we have defined
the function
\eq{
  \label{def_zeta}
  \vartheta(\xi) =  \frac{1}{1+\bigl( \frac{R_2}{R_1} f \xi\bigr)^2} \,.
}  
The non-vanishing components of the dual field strength, determined via the equations in 
\eqref{buscher_02}, read
\eq{
  \label{ex3_t-h}
  \mathcal H_{\eta \xi\chi} = 2 \sin(2\eta)\op\vartheta(\xi) \,,\hspace{45pt}
  \mathcal H_{X\xi \chi} = -f \left(\frac{R_2}{R_1}\right)^2  \vartheta^2(\xi) \op\Bigl[ 1- \bigl( \tfrac{R_2}{R_1} \op f\op \xi \bigr)
  ^2
  \Bigr] \,.
}
Note that this background is locally geometric, but is globally not well-defined. 
Indeed, when going around the circle
in the $\xi$-direction as $\xi\to\xi+2\pi$,  the metric given by \eqref{ex3_t-geom} and the components 
of the field strength \eqref{ex3_t-h} are not periodic. The mismatch cannot be compensated 
by having diffeomorphism as transition functions between different charts, since the 
former is due to the monodromy  \eqref{ex3_monodd} of the Killing 
vector $\mathsf K_{(2)}$.
However, following the same reasoning as illustrated in figure~\ref{fig_05}, the dual space can be 
interpreted as a T-fold.


\subsection{T-fold}

It is beyond the scope of this paper to analyze the above T-fold background in further detail,
and we refer this question to a later point. Nevertheless, let us 
make the following remarks. 
In order to simplify the metric in \eqref{ex3_t-geom}, we introduce as basis of one-forms 
as follows
\eq{
  &\mathbb E^{\xi} = d\xi \,,  \\
  &\mathbb E^{\eta} = d\eta \,, \hspace{60pt} \mathbb E^{X} = dX - f\cos(2\eta) \op\xi \op d\xi \,, \\[2pt]
  &\mathbb E^{\chi} = d\chi \,,  
}
where in particular $\mathbb E^X$ is not single-valued and not closed
\eq{
  d\mathbb E^{\xi} = d\mathbb E^{\eta} =  d\mathbb E^{\chi} = 0 \,, \hspace{60pt}  
  d\mathbb E^{X} = 2 f\sin(2\eta) \op\xi \op d\eta\wedge d\xi \,.
}  
Employing this basis, the metric of the T-fold can be expressed via the following line-element squared as
\eq{
  \label{ex5_t_fold}
  ds^2 =
  R_1^2 \Bigl[  \op\bigl(\mathbb E^{\eta}\bigr)^2 + \sin^2(2\eta) \bigl(\mathbb E^{\xi}\bigr)^2 \op\Bigr] +
  \vartheta(\xi)
  \left[\,
  \tfrac{1}{R_1^2}\,\bigl(\mathbb E^{\chi}\bigr)^2 + R_2^2 \op\bigl( \mathbb E^X \bigr)^2 \right] 
  ,
}
where the function $\vartheta(\xi)$ was defined in equation \eqref{def_zeta}.
The line element \eqref{ex5_t_fold} describes  a {\em local} two-torus $\widetilde{\mathbb T}^2$ along $\mathbb E^{\chi}$ and
$\mathbb E^X$, which is non-trivially fibered over a two-sphere. 
We also remark that for a vanishing twisting $f=0$, we obtain the geometric background $S^2\times \mathbb T^2$.


\subsubsection*{Chain of T-duality transformations}

Let us finally summarize the chain
of T-duality transformations studied in this section.
After slightly adjusting our notation and denoting by $\mathcal E_h$ and $\widetilde{\mathcal E}_h$ the 
global and local circle bundles with twisting $h$ discussed above, we arrive at the following picture:
\eq{
  \begin{array}{c@{\hspace{10pt}}||r@{\hspace{4pt}}c@{\hspace{-22pt}}c@{\hspace{-22pt}}c@{\hspace{5pt}}l@{\hspace{35pt}}l}
  \mbox{background} & \multicolumn{5}{c}{\mbox{geometry}\hspace{15pt}} & \mbox{field strength} \\
  \hline\hline 
  && \\[-5pt]
  \mbox{$H$-flux}
  &&&\hspace{7pt}
  S^1_{X} \:\times\: S^3_{\xi_1,\xi_2,\eta} 
  \hspace{-7pt}
  &&& H = h\, d\xi_1 \wedge d\xi_2 \wedge dX \\
  &&&\hspace{6pt}
  \rotatebox{270}{$\displaystyle   \xleftrightarrow{\quad \rotatebox{90}{$\scriptstyle\mathsf T_X$}\quad}$} 
  \hspace{-6pt}
  \\[43pt]
  \mbox{$f$-flux}
  &S^1_X  & \hookrightarrow &\mathcal E_h &\rightarrow & S^3_{\xi_1,\xi_2,\eta} & H=0 \\
  &&&\hspace{6.5pt}
  \rotatebox{270}{$\displaystyle  \xleftrightarrow{\quad \rotatebox{90}{$\scriptstyle\mathsf T_{\xi_1}$}\quad}$}
  \hspace{-6.5pt}  
  \\[43pt]
  \mbox{$Q$-flux}
  &\hspace{10pt}\widetilde{\mathbb T}_{X,\xi_1}& \hookrightarrow &\widetilde{\mathcal E}_h &\rightarrow &
  S^2_{\xi_2,\eta} & \widetilde H\neq0  \\
  &&&\hspace{6.5pt}
  \rotatebox{270}{$\displaystyle  \xleftrightarrow{\quad \rotatebox{90}{$\scriptstyle\mathsf T_{\xi_2}$}\quad}$}
  \hspace{-6.5pt}  
  \\[43pt]
  \mbox{$R$-flux}
  &&& \hspace{-5pt}\ldots \\[-8pt]
  &&&&&&\textcolor{white}{\cdot}
  \end{array}
}
Given these relations, it is then tempting to speculate that a further T-duality transformation on the 
T-fold  gives rise to an $R$-flux background with non-as\-so\-cia\-tive features.
However, as we mentioned above, this discussion is beyond the scope of this paper.


\section{Summary and conclusions}
\label{sec_sc}

In this paper, we have reviewed the  transformation rules of the 
metric, Kalb-Ramond field and dilaton under T-duality. However, instead of expressing
the formulas in terms of the Kalb-Ramond field $B$ itself, as it is usually done for the Buscher rules,
we described the T-dual background employing the corresponding field strength $H=dB$.
In sections~\ref{sec_tori} and~\ref{sec_spheres} we have then illustrated 
our formalism with a detailed discussion of T-duality transformations for tori and  spheres.

The Buscher rules have long been studied in the literature from different perspectives,
and are rather well-understood. Nevertheless, in this paper we  were able derive
some novel results and provide new interpretations on this subject.
\begin{itemize}

\item In particular, in section~\ref{sec_sigma} we reviewed the sigma-model action for the closed string
for a non-vanishing field strength $H$ of the Kalb-Ramond field. We found that 
the symmetry structure of this action can be described via the $H$-twisted Courant bracket, which 
agrees with similar results  in \cite{Alekseev:2004np} obtained in a different context.

\item In section~\ref{sec_t-duality} we derived the transformation rules of the metric,
Kalb-Ramond field and dilaton under T-duality. This was done through gauging the sigma-model action
by a target-space isometry. We then observed that 
the remaining global symmetries of the gauged action are determined by \eqref{constr_02}. This
explains how under T-duality the isometry group can be reduced, which we illustrated with 
examples in sections~\ref{sec_tfold} and~\ref{sec_sphere}.

\item In the course of the derivation of the T-duality rules, we made us of an enlarged 
target space \cite{Rocek:1991ps,Alvarez:1993qi}, for which a metric $\check G$ and field strength $\check H$ 
can be defined. 
We  noted that the metric $\check G$ has a null-eigenvector, which can be used to obtain
a convenient set of coordinates leading to a dual target-space background.
However, this can only be done consistently if the constraint \eqref{coc_04} is met.
For many examples \eqref{coc_04} is automatically satisfied, but we believe 
that this restriction has not appeared in the literature until now.

\item In contrast to the Buscher rules given for the metric $G$ and Kalb-Ramond field $B$,
here we expressed the T-duality transformation rules in terms of $G$ and the field strength 
$H=dB$.\op\footnote{T-duality transformation rules involving the field strength $H$ have also appeared in 
\cite{Bandos:2003bz}, but without a detailed derivation from a world-sheet point of view.}
This has the advantage that we do not have to rely on a choice of gauge for the initial configuration.
For the dual background, the topology is  specified by the $H$-flux, and for the geometry 
there is a freedom of choosing local coordinates. 
This is in accordance with the Buscher rules, where a choice of gauge for $B$ 
determines the dual geometry.

\item In section~\ref{sec_tori} we have illustrated the above-mentioned transformation rules with the example 
of the three-torus. The results obtained in our formalism agree with those known in the literature,
however,  we were able to generalize these findings by allowing for instance for 
general T-duality directions. We furthermore discussed possible monodromies of
Killing vectors, and in figure~\ref{fig_05} we interpreted the T-fold from a
point of view which does not involve a doubled geometry.

\item Finally, in section~\ref{sec_spheres} we studied T-duality transformations for spheres.
We first reviewed and generalized the discussion in \cite{Alvarez:1993qi} about the three-sphere,
and then constructed a new example of a non-geometric background.

\end{itemize}

The results obtained in this paper motivate further studies in this direction.
First, it would be interesting to study the T-fold based on the sphere as a supergravity background,
and investigate whether conclusions similar to those in \cite{Shelton:2005cf} can be drawn.
Furthermore, the approach to analyze T-duality transformations presented here might 
be suitable to find new non-geometric backgrounds which are not based on the torus. 
Second, since the sigma model on a three-sphere with $H$-flux corresponds to the $SU(2)$ WZW model, which is 
conformal, it might be possible to investigate the T-fold of section~\ref{sec_spheres} 
as a proper string-theory background. This could lead to a better string-theoretical understanding of
non-geometric spaces. We hope to return to these questions in the future.


\vskip2cm
\subsubsection*{Acknowledgements}

We would like to thank Ralph Blumenhagen, Gianguido Dall'Agata, Luca Martucci and 
especially Felix Rennecke for very helpful discussions, furthermore Gianguido Dall'Agata for
useful comments on the manuscript, and Larisa Jonke for correspondence on the Jacobiator of 
the $H$-twisted Courant bracket.
We also thank the referee at JHEP for helpful comments.
The author is supported 
by the MIUR grants PRIN 2009-KHZKRX and FIRB RBFR10QS5J.


\clearpage
\bibliography{references}  
\bibliographystyle{utphys}


\end{document}